\documentclass[aps,pra,twocolumn,superscriptaddress,nofootinbib,longbibliography]{revtex4-1}
\usepackage{amsmath}
\usepackage{amscd}
\usepackage{wasysym}
\usepackage{hyperref}
\usepackage{color}
\usepackage{braket}
\usepackage{float}
\usepackage[T1]{fontenc}
\usepackage{ae,aecompl}
\usepackage{amsfonts}
\usepackage{amsthm}
\usepackage{dsfont}
\usepackage{graphicx}
\usepackage{amsfonts}
\usepackage{physics}
\graphicspath{{./figs/}}

\newcommand{\ue}{\mathrm{e}}

\newcommand{\la}{\langle}
\newcommand{\ra}{\rangle}

\newcommand{\rh}{\hat{\rho}}
\newcommand{\sh}{\hat{\sigma}}
\newcommand{\Mh}{\hat{M}}

\newcommand{\R}{\mathbb{R}}

\newcommand{\figref}[1]{Fig.~\ref{#1}}

\begin{document}

\title{Steering-based randomness certification with squeezed states and homodyne measurements}

\author{Marie Ioannou\textsuperscript{*}}%\thanks{These authors contributed equally to this work}
\address{Department of Applied Physics, University of Geneva, 1211 Geneva, Switzerland}

\author{Bradley Longstaff\textsuperscript{*}}%\thanks{These authors contributed equally to this work}
\address{Center for Macroscopic Quantum States (bigQ), Department of Physics, Technical University of Denmark, Fysikvej, 2800 Kgs. Lyngby, Denmark}

\author{Mikkel V. Larsen}
\address{Center for Macroscopic Quantum States (bigQ), Department of Physics, Technical University of Denmark, Fysikvej, 2800 Kgs. Lyngby, Denmark}

\author{Jonas S. Neergaard-Nielsen}
\address{Center for Macroscopic Quantum States (bigQ), Department of Physics, Technical University of Denmark, Fysikvej, 2800 Kgs. Lyngby, Denmark}

\author{Ulrik L. Andersen}
\address{Center for Macroscopic Quantum States (bigQ), Department of Physics, Technical University of Denmark, Fysikvej, 2800 Kgs. Lyngby, Denmark}

\author{Daniel Cavalcanti}
\address{Bitflow, C/ Piquer 23, 08004 Barcelona, Spain}

\author{Nicolas Brunner}
\address{Department of Applied Physics, University of Geneva, 1211 Geneva, Switzerland}

\author{Jonatan Bohr Brask}
\address{Center for Macroscopic Quantum States (bigQ), Department of Physics, Technical University of Denmark, Fysikvej, 2800 Kgs. Lyngby, Denmark}

\begin{abstract}
High-quality randomness, certified to be unpredictable by eavesdroppers, is key to secure information processing. Quantum mechanics enables randomness certification with minimal trust in the devices used, by exploiting quantum nonlocality. However, such full device independence is challenging to implement. We present a scheme for quantum randomness certification based on quantum steering. The protocol is one-sided device independent, providing high security, but requires only states and measurements that are simple to realise on quantum optics platforms -- squeezed vacuum states and homodyne detection. This ease of implementation is demonstrated experimentally and implies that giga-hertz random bit rates should be attainable with current technology. Furthermore,  our scheme is immune to the detection loophole and represents the closest to full device independence that can be achieved using purely Gaussian states and measurements.
\end{abstract}

\maketitle

\def\thefootnote{*}\footnotetext{These authors contributed equally to this work.} \def\thefootnote{\arabic{footnote}}\setcounter{footnote}{0}

Randomness is an important resource in science and technology for simulations and information processing. In particular, random numbers that are unpredictable by any adversary are key to cryptographic security \cite{Hayes2001}. Random numbers can be generated from hard-to-predict physical processes, and pseudo-random-number generators, implemented in software, can expand short random seeds into longer sequences that appear random. However, classical physics is fundamentally deterministic, as are software algorithms. Therefore, guaranteeing security based on classical random-number generation requires assumptions about the knowledge and computational resources available to potential eavesdroppers. Such assumptions may be difficult to justify as the adversaries might not be known.

Randomness generation based on quantum physics provides an alternative free of this limitation \cite{Acin2016,Herrero2017,Bera2017}. For quantum systems, there exist measurements whose outcomes cannot be predicted even given a complete quantum-mechanical description of the system and measurement device. This implies that security can be guaranteed based only on the user's own knowledge, as long as the adversary is bound by quantum mechanics. That is, the user needs only trust their own characterisation of the randomness-generation device. For example, randomness can be generated by detecting the output path of a single photon impinging on a beam splitter \cite{Stefanov2000}. When the beam splitting ratio and other characteristics of the setup are known, the unpredictability of the outcome can be certified relative to any quantum adversary, regardless of their computational power or available resources.

Remarkably, exploiting the nonlocality \cite{Bell1964,Brunner2014} of quantum mechanics allows randomness certification even with almost uncharacterised devices. In setups violating a Bell inequality, randomness can be certified device independently, i.e., without making any assumptions about the inner workings of the devices used \cite{colbeckPhD2009,Pironio2010}. This represents a very strong form of security, as the devices can be largely untrusted, and it has been demonstrated in several experiments \cite{Pironio2010,Christensen2013,Bierhorst2018,Liu2018,Shalm2021,Liu2021}. However, it is also more challenging to implement than the device-dependent approach because loophole-free Bell violation requires low noise and high detection efficiencies. This motivates the search for trade-offs, where full device independence is relaxed in order to gain ease of implementation, while still maintaining high security. Many works have explored this semi-device-independent setting in prepare-and-measure setups without nonlocality, by allowing source or measurement devices to be partially characterised, see e.g.~\cite{Li2011,Vallone2014,Lunghi2015,Mironowicz2021,Cao2015,Marangon2017,Cao2016,Xu2016,Brask2017a,Gehring2021,Michel2019,Rusca2019,Drahi2020,Rusca2020,Avesani2021}. An alternative approach is to exploit Einstein-Podolsky-Rosen steering \cite{Einstein1935,Reid1989,Wiseman2007}, which is a form of nonlocality intermediate between full Bell nonlocality and quantum entanglement. In a bipartite steering scenario, the device of one party is untrusted while that of the other party is characterised. This setting is thus one-sided device independent, and has been considered for applications in quantum cryptography \cite{Reid2000,Branciard2012} and randomness generation \cite{Law2014,Passaro2015,Skrz2018}. While experiments on quantum key distribution were demonstrated \cite{Gehring2015,Walk2016}, the practical implementation of these ideas for QRNG is mostly unexplored \cite{Mattar2017,Wang2018}.

\begin{figure*}[t]
    \centering
		\includegraphics[width=0.98\textwidth]{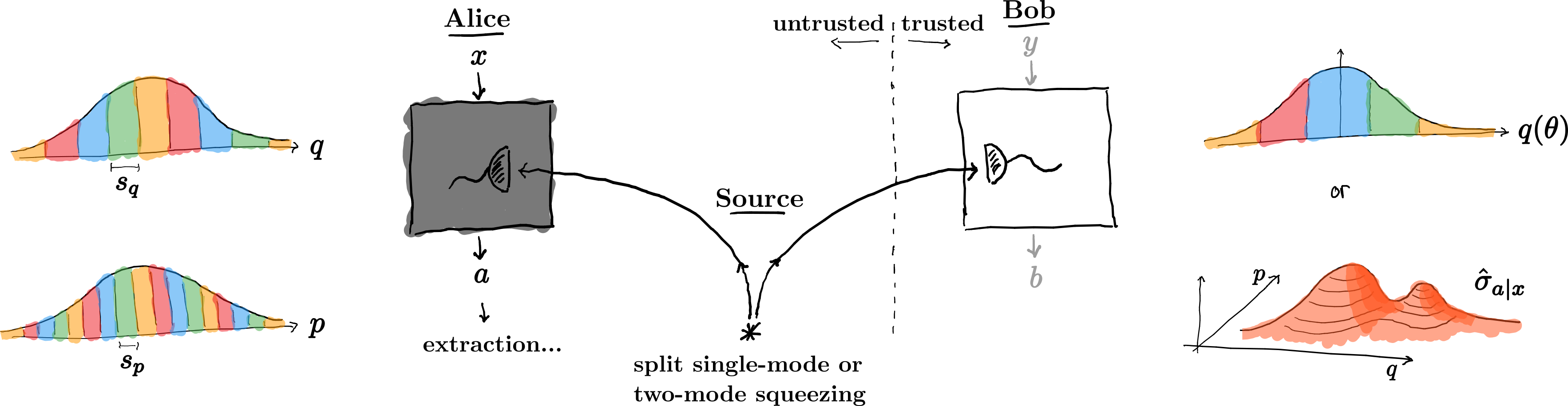}
	\caption{Setup for steering test and randomness certification using Gaussian states and measurements, consisting of a source of squeezed, entangled states and two parties, Alice and Bob, who perform homodyne measurements. The source emits either two-mode squeezed vacuum or single-mode squeezeed vacuum split on a balanced beam splitter. Alice measures one of two conjugate quadratures, according to a binary input $x$, and applies periodic binnings into $o_A$ outcomes $a$. Bob either performs full state tomography or measures $m_B$ different quadratures, according to input $y$, and applies a non-periodic binning into $o_B$ outcomes $b$. For certifying steering and randomness, Alice's device and the source state are untrusted, while Bob's device is assumed to be well characterised. Randomness is extracted from Alice's measurement outcome, for one of her inputs.}
	\label{fig.concept}
\end{figure*}

Here, we develop a steering-based quantum randomness-certification protocol that can be implemented with simple light sources and measurements. The setup requires only squeezed light and homodyne detection, and can tolerate realistic levels of loss and noise. It is thus readily implementable with existing technology, which we demonstrate by applying our protocol to data from the experiment of Ref.~\cite{Larsen19}. Randomness is certified, and we estimate that a rate of 70 kbits/s could be extracted. In a dedicated setup, significantly higher rates are expected. Fast squeezing sources, operating in the THz range, and homodyne detection in the GHz range have been realised \cite{Kashiwazaki2020}. Combined with a higher entropy per round, this should enable secret bit rates in the GHz range. We note that the scheme is free of any detection loophole, because (unlike single-photon detection) homodyne detection always provides an output and no data is discarded. Furthermore, the setup uses only Gaussian states and measurements. Bell nonlocality, and hence full device independence, is impossible with only Gaussian resources (this follows from positivity of the Gaussian Wigner functions and Fine's theorem \cite{Fine1982}). Thus, our protocol in this sense provides the closest to device independence one may hope for in this setting.

Our work exploits entangled squeezed states, which are infinite dimensional, and homodyne measurements, which have continuous outcomes. Steering has been demonstrated with such resources \cite{Ou1992,Haendchen2012,Armstrong2015,Deng2017,Qin2017,Wang2020}. However, for quantifying randomness it is convenient to work with measurements with a finite number of outcomes, where powerful methods based on semidefinite programming can be applied \cite{Cavalcanti2016}. This can be achieved by coarse-graining the homodyne outcomes into a finite number of bins. To guide the choice of binning, we note that, as the dimension grows, a bipartite maximally entangled state in finite dimension approaches a two-mode infinitely squeezed vacuum state. In finite dimensions, Skrzypczyk and Cavalcanti \cite{Skrz2018} found that optimal steering-based randomness generation is achieved by performing mutually unbiased measurements on maximally entangled states. The optimal measurements are conjugate, i.e.~related by a Fourier transform. This suggests that randomness can be obtained by measurements of conjugate quadratures on two-mode squeezed states. In Ref.~\cite{Tasca2018}, Tasca et al. identified coarse grainings of homodyne measurements that preserve mutual unbiasedness. One may therefore expect that adopting this binning scheme will enable steering and randomness certification even at finite squeezing. Our results confirm this intuition.
\begin{figure*}[t]
		\includegraphics[width=\textwidth]{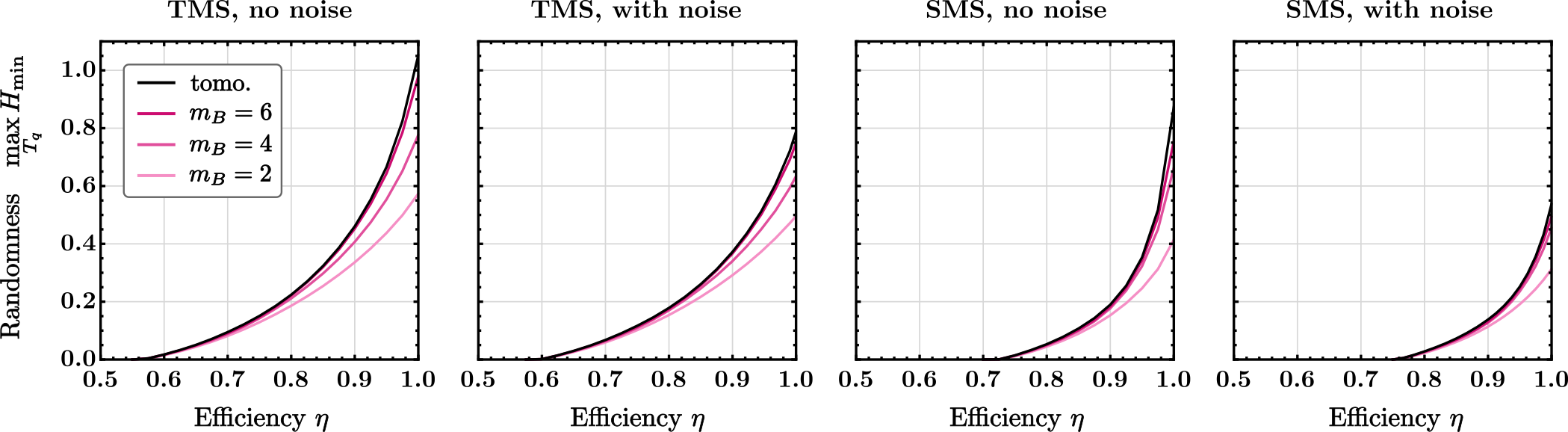}
	\caption{Optimal min-entropy vs.~transmission efficiency, for the two-mode squeezed (TMS) and the split single-mode squeezed (SMS) vacuum source with and without noise (0.01 shot-noise units). $H_{\textnormal{min}}$ is maximised over Alice's binning period in the range $T_q \in [2,10]$. In all cases Alice has a choice of $m_A = 2$ observables, $q$ and $p$, binned to $o_A = 8$ outcomes. Bob performs tomography (black curve) or measures in $m_B=2,4,6$ different directions equally spaced between $q$ and $p$, binned to $o_B = 16$ outcomes. For TMS the squeezing is $-4$ dB, and for SMS it is $-6$ dB.}
	\label{fig.Hmin_eta}
\end{figure*}

We consider a bipartite setup, as illustrated in \figref{fig.concept}. An entangled state $\rh$ is distributed to two parties Alice and Bob. For the purpose of certifying steering and randomness, Alice and the source are untrusted, while Bob's device is well characterised. In each round, Alice chooses one of $m_A$ measurements each with a number $o_A$ of outcomes. We denote her input (choice) $x$ and output (outcome) $a$. Thus $x$ and $a$ can take $m_A$ and $o_A$ different values, respectively. Bob either performs full state tomography or some fixed set of measurements. In the former case, the information available after many repetitions consists of the conditional input-output probabilities of Alice $P(a|x)$ and the conditional states of Bob $\rh_{a|x}$, or equivalently, in the assemblage of unnormalised states $\sh_{a|x} = P(a|x) \rh_{a|x}$. A precise definition of steering was given in \cite{Wiseman2007}. An assemblage is said to be steerable if it does not admit a local-hidden-state (LHS) model
\begin{equation}
\sh_{a|x} = \int_\Lambda \pi(\lambda)P(a|x,\lambda)\rh_\lambda\,d\lambda ,
\end{equation}
where $\pi(\lambda)$ is a probability distribution over $\lambda$, which can be thought of as a classical common cause that determines both the output $a$ and the quantum state $\rh_\lambda$ of Bob. A steerable assemblage cannot be explained in terms of a classical common cause, and in particular $\rh$ must then be entangled. Furthermore, a key observation from the point of view of randomness certification is that the lack of a LHS model implies that one cannot have $P(a|x) \in \{0,1\}$ for all $a,x$, i.e.~$P(a|x)$ cannot be completely deterministic \cite{Passaro2015}. In other words, if Bob's assemblage is steerable then there must be some randomness in Alice's measurement outcomes. Note that randomness is extracted from the untrusted party and that only steering from Alice to Bob is required.

The amount of certifiable randomness can be quantified in terms of the maximal probability for an eavesdropper (Eve) to correctly predict the output given knowledge of the input and other available side information (in particular, we allow Eve to be entangled with the source, but we do assume rounds to be independent and identically distributed with respect to Eve). We consider randomness to be generated for a particular input $x^*$ and denote the corresponding guessing probability $p_g(x^*)$. By the leftover hash lemma~\cite{Impagliazzo1989}, the asymptotic number of almost uniformly random bits extractable per round is given by the min-entropy $H_{\textnormal{min}}(x^*) = -\log_2 p_g(x^*)$. The guessing probability can be computed via the following optimisation problem \cite{Passaro2015,Cavalcanti2016}
\begin{subequations}
\label{eq.pgSDPtomo}
\begin{align}
\max_{\{\hat{\sigma}_{a|x}^e\}} & \quad\!\! \tr\left[\sum_e \hat{\sigma}_{a=e|x^\ast}^e\right], \\
\label{eq.pgSDPobs}
\textnormal{s.t.}\quad &\sum_e \hat{\sigma}_{a|x}^e = \hat{\sigma}_{a|x}^{\textnormal{obs}} \qquad\quad \forall a,x,\\
\label{eq.pgSDPnosig}
&\sum_a \hat{\sigma}_{a|x}^e = \sum_a \hat{\sigma}_{a|x'}^e \quad\: \forall e,x \neq x',\\
\label{eq.SDPpos}
&\qquad\! \hat{\sigma}_{a|x}^e \geq 0 \qquad\qquad\; \forall a,x,e .
\end{align}
\end{subequations}
This is equivalent to optimising over all strategies of Eve that are compatible with the observed assemblage $\sh_{a|x}^{\textnormal{obs}}$ \eqref{eq.pgSDPobs} and with no-signalling from Alice to Bob and Eve \eqref{eq.pgSDPnosig} \cite{Passaro2015}. Note that \eqref{eq.pgSDPtomo} is a semidefinite program (SDP) and can be solved efficiently numerically~\cite{Boyd2004}.

Performing full state tomography can be demanding experimentally, and it is then desirable to restrict Bob to some, ideally small, set of $m_B$ measurements with $o_B$ outcomes. In this case, the available observation from the experiment is not the assemblage but the conditional probabilities $P(ab|xy)$, where $y$ and $b$ label Bob's input and output, respectively. Randomness can still be certified and the guessing probability can again be computed via an SDP. Assuming that Bob performs positive-operator-valued measures (POVMs) with elements $\Mh_{b|y}$, the guessing probability is again given by the optimisation \eqref{eq.pgSDPtomo}, except that the condition \eqref{eq.pgSDPobs} is replaced by the requirement that Eve's strategy must reproduce the observed probabilities, $\sum_e \tr\left[\Mh_{b|y}\sh_{a|x}^e\right] = P(ab|xy) \quad \forall a,b,x,y$.

We now determine the amount of randomness certifiable in a setup using squeezed light and homodyne detection. The source distributes either a two-mode squeezed (TMS) vacuum state or a single-mode squeezed (SMS) vacuum state split on a balanced beam splitter (see \figref{fig.concept}). We let $q_A$, $p_A$ and $q_B$, $p_B$ denote conjugate quadratures for Alice and Bob respectively. The initial states are chosen such that in the split single-mode case, $q_A + q_B$  is squeezed, and in the two-mode case, both $q_A + q_B$ and $p_A-p_B$ are squeezed. Alice makes $m_A=2$ measurements of $q_A$ and $p_A$ (note that the local oscillator required for homodyne detection does not open up any loophole as it is untrusted). Following Ref.~\cite{Tasca2018}, her results are binned into $o_A$ outcomes, resulting in POVMs
\begin{equation}
\label{eq.binnedM}
\hat{M}_{a|x} = \int_\R f_a(z,T_x) \ket{z}_x\!\bra{z} dz ,
\end{equation}
where $x=q,p$ is the input, $\ket{z}_x$ are $x$-quadrature eigenstates, and $f_a(z,T_x)$ is a periodic mask function
\begin{equation}
f_a(z,T_x) =
\begin{cases}
1, & as_x \leq z \mkern-12mu\mod T_x < (a+1)s_x, \\
0, & \text{otherwise}.
\end{cases}
\end{equation}
Here,  $T_x$ is the period, $s_x = T_x/o_A$ the width of the bins (see \figref{fig.concept}), and $T_p = 2\pi / s_q$ to ensure mutual unbiasedness. We take $a \in \{0,\ldots,o_A-1\}$.

Bob either performs tomography or a fixed set of measurements. In principle, optimal measurements could be determined (at least numerically) from the dual of the SDP \eqref{eq.pgSDPtomo} for tomography which provides an optimal steering inequality. However, it is not clear that these measurements can be realised in practice, or how they might be approximated. Instead, we let Bob perform binned homodyne measurements as well. Specifically, $m_B$ quadrature measurements along directions in phase space equally spaced between $q_B$ and $p_B$. He applies a binning consisting of $o_B-1$ intervals dividing the range $[-r,r]$ evenly, and the last bin constitutes everything outside this range. We found that setting $r = 5\sigma$, where $\sigma^2$ is the largest variance in the (Gaussian) entangled initial state (i.e.~the largest diagonal entry of the covariance matrix), works well for our parameter values.  The central binned region is then sufficiently wide to capture the variation induced by Alice's measurements while also admitting sufficiently narrow bins for Bob's outcomes to reveal this variation without $o_B$ being intractably large.

We model detector inefficiencies and other losses by fictitious beam splitters with transmittivity $\eta$ between the source and each party. We take the losses to be symmetric for Alice and Bob, and we consider both pure loss, with vacuum entering the other port of the beam splitters, and noise, modeled by replacing the vacuum with thermal states. We compute the observed data ($\sh_{a|x}$ or $P(ab|xy)$) starting from the covariance matrix of the joint Gaussian state, including loss and noise. A derivation of the covariance matrix is provided in App.~\ref{appendix:gaussian states}. In order to implement the SDPs for the guessing probability, we need to work in finite dimension. We therefore calculate the Fock-space representation of the state and measurement operators, applying a cut off in photon number, and compute the data from there. The cut off is chosen sufficiently large to not affect the final results, see App.~\ref{appendix:finite dimension}. We then run the SDPs given above to determine the guessing probability and min-entropy in each case. Finally, we optimise over Alice's binning period $T_q$. 

The results are summarised in \figref{fig.Hmin_eta}. We observe several interesting features. First, randomness can be generated at moderate levels of squeezing, with results shown for $-4\,\text{dB}$ for the TMS source and $-6\,\text{dB}$ for split SMS. Second, a significant amount of randomness can be certified even for sizable loss and the entropy is non-zero above $\eta \gtrsim 0.55$ for the TMS source and $\eta \gtrsim 0.75$ for split SMS. Third, allowing for added noise corresponding to 1\% of the vacuum variance (0.01 shot-noise units, see App.~\ref{appendix:gaussian states}) does not dramatically decrease the performance. These numbers indicate that implementation of our protocol is well within reach of contemporary experimental techniques. Finally, performing just a few binned homodyne measurements for Bob is almost as good as tomography. For $m_B=6$ measurements, $H_{\textnormal{min}}$ is within a few percent of the full-tomography result, and with just $m_B=2$ one obtains about half of the optimal entropy. This shows that the protocol already performs well in the simplest setting of just two measurements per party.

Indeed, we can provide a proof-of-principle demonstration of the practicality of the protocol by applying it to existing experimental data, showing that randomness can in fact be certified in a setup that has already been realised. In Ref.~\cite{Larsen19}, Larsen et al.\ implemented a two-mode squeezed vacuum source by temporal multiplexing in fibre and characterised it via homodyne measurements of the two output modes. Assigning the two modes to Alice and Bob, respectively, an appropriate subset of the characterisation measurements corresponds to $q$,$p$-quadrature for each party, i.e.~to the case of two settings per party, $m_A = m_B = 2$. Post-processing the data, we can then apply binnings according to the strategies outlined above and estimate the joint probabilities $P(ab|xy)$. For Bob's binning, we use $r = 5$. Each data set (for a given combination of quadratures) contains 16,000 measurements, and we calculate $P_{\textnormal{exp}}(ab|xy)$ from the frequencies of the outcomes.

Owing to finite statistics, the distribution $P_{\textnormal{exp}}(ab|xy)$ is signalling, and hence cannot be used directly as a constraint in the SDP for computing $H_{\textnormal{min}}$ (because the SDP is then always infeasible as the distribution cannot be obtained from any quantum strategy for Eve). Instead, we construct an idealised theoretical model of the experiment and obtain an approximation of the initial Gaussian state $\hat{\rho}_G$. We then compute the probability distributions $P_\textnormal{theory}(ab|xy)=\textnormal{tr}[\hat{M}_{a|x}\otimes\hat{M}_{b|y}\hat{\rho}_G]$, which are guaranteed to be no-signalling, and use these in the SDP. Finally, we extract the corresponding dual variables and use them together with the experimental distributions $P_\textnormal{exp}(ab|xy)$ to obtain a lower bound on the min-entropy of the experimental data (see App.~\ref{appendix:lower bound} for details).

\begin{figure}[t]
	\begin{center}
		\includegraphics[width=0.35\textwidth]{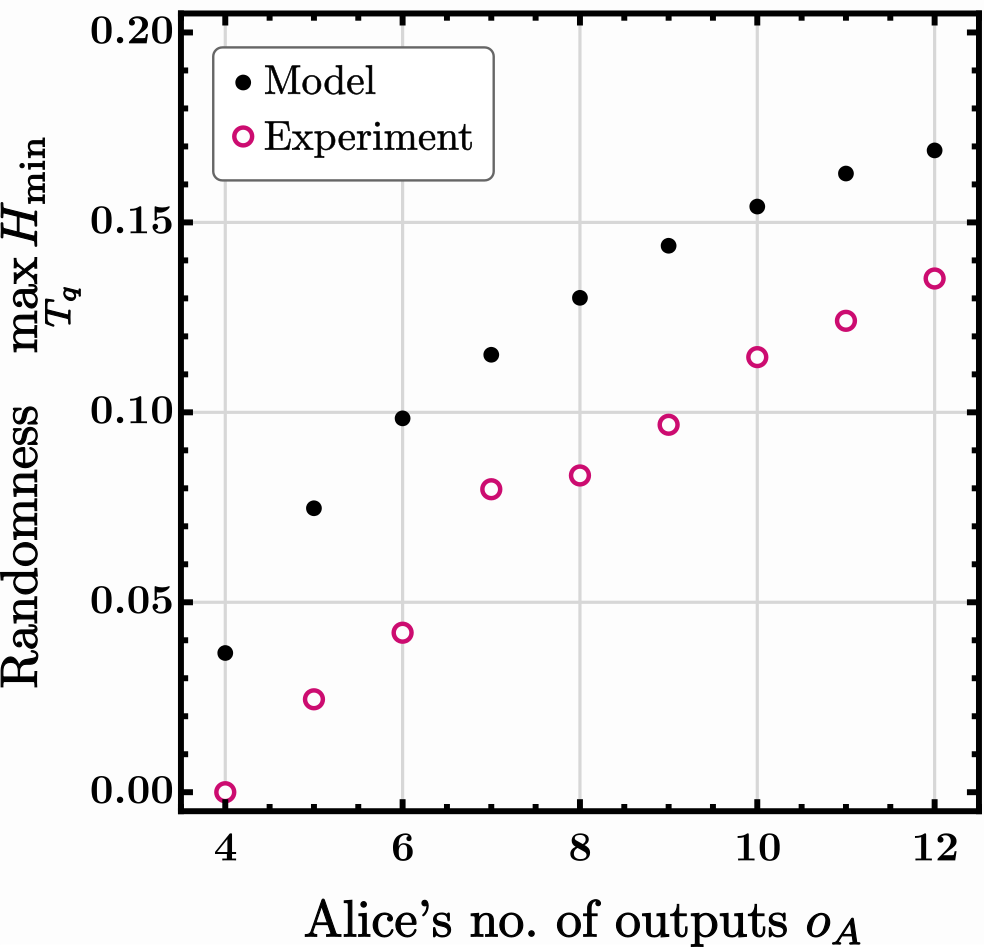}
	\end{center}
	\caption{Results obtained from experimental data of Ref.~\cite{Larsen19}. Min-entropy vs.~Alice's number of outputs. The optimal values of $H_{\textnormal{min}}$ from the idealised theoretical model (black dots) and the lower bounds on $H_{\textnormal{min}}$ from the experimental data (pink circles) are pictured. Alice and Bob both measure $q$ and $p$, and randomness is extracted from Alice's $q$-measurement. $H_{\textnormal{min}}$ is maximised over Alice's binning period in the range $T_q \in [2,10]$. Bob has $o_B = 16$ outcomes.}
	\label{fig.exp}
\end{figure}

The resulting optimal min-entropy is shown in \figref{fig.exp} as a function of the number of outputs for Alice. We see that the experimentally certified lower bound on the min-entropy is in good agreement with the idealised theoretical model. The model predicts that about 0.17 bits of randomness per round can be certified with $o_A=12$ and $o_B=16$, with a lower bound of about 0.14 bits of randomness per round. While $H_{\textnormal{min}}$ might increase further, for computational reasons we cannot employ larger numbers of outputs. The observed squeezing (in the relevant temporal mode) is $-3.88$ dB in $q_A + q_B$ and $-3.76$ dB in $p_A - p_B$, and the overall efficiency is 68\%. Furthermore, the repetition rate of the experiment was 500 kHz from which we get an approximate extracted random bit rate of $\sim 70$ kbits/s. These results clearly show that our scheme is feasible in practice. We expect that significantly higher $H_{\textnormal{min}}$ could be attained in a dedicated experiment. In particular, it should be possible to significantly improve the overall efficiency to around 90\% and to lower phase noise, thus improving the squeezing level, by avoiding the use of optical switching and fiber delays. Also, fast sources and detectors should enable GHz repetition rates \cite{Kashiwazaki2020}, leading to GHz-range secret bit rates.

In conclusion, we presented a scheme for quantum random-number certification at the one-sided device-independent security level which can be realised using purely Gaussian resources, namely squeezed states and homodyne detection. The scheme is robust to realistic levels of loss and noise and can certify significant amount of randomness (min-entropy approaching 1) for moderate squeezing levels well below 10 dB. It is hence feasible to implement with standard technology, as we also have shown by applying the protocol to existing experimental data from Ref.~\cite{Larsen19}, providing a proof of principle.

One interesting future direction would be an experiment designed specifically for this protocol, which would likely perform significantly better. Spatial separation of the parties could also be implemented. A more thorough analysis accounting for finite-size effects would be required for an accurate calculation of the entropy. In particular, to certify more randomness than consumed (i.e.\ to achieve randomness expansion), the inputs should be biased, with $x^*$ occurring more often while allowing $P(ab|xy)$ to be estimated sufficiently well. This trade-off can be made rigorous in a finite-size analysis. Ideally real-time randomness extraction should also be applied.

We note that a complementary work demonstrating steering-based randomness certification with discrete variables appeared simultaneously with this paper \cite{Joch2021}.

\emph{Ackowledgements.} JBB and BL acknowledge support from the Carlsberg Foundation and the Independent Research Fund Denmark 7027-00044B. MI and NB acknowledge funding from the EU Q Flagship project QRANGE and the Swiss National Science Foundation (project 2000021\_192244/1 and NCCR QSIT). MVL, JSN, and ULA acknowledge the Danish National Research Foundation through the Center for Macroscopic Quantum States (bigQ, DNRF0142) and the EU’s Horizon 2020 research and innovation programme under grant agreement No 820466 (CiViQ).

\bibliography{qrng_steering}

\begin{thebibliography}{58}%
\makeatletter
\providecommand \@ifxundefined [1]{%
 \@ifx{#1\undefined}
}%
\providecommand \@ifnum [1]{%
 \ifnum #1\expandafter \@firstoftwo
 \else \expandafter \@secondoftwo
 \fi
}%
\providecommand \@ifx [1]{%
 \ifx #1\expandafter \@firstoftwo
 \else \expandafter \@secondoftwo
 \fi
}%
\providecommand \natexlab [1]{#1}%
\providecommand \enquote  [1]{``#1''}%
\providecommand \bibnamefont  [1]{#1}%
\providecommand \bibfnamefont [1]{#1}%
\providecommand \citenamefont [1]{#1}%
\providecommand \href@noop [0]{\@secondoftwo}%
\providecommand \href [0]{\begingroup \@sanitize@url \@href}%
\providecommand \@href[1]{\@@startlink{#1}\@@href}%
\providecommand \@@href[1]{\endgroup#1\@@endlink}%
\providecommand \@sanitize@url [0]{\catcode `\\12\catcode `\$12\catcode
  `\&12\catcode `\#12\catcode `\^12\catcode `\_12\catcode `\%12\relax}%
\providecommand \@@startlink[1]{}%
\providecommand \@@endlink[0]{}%
\providecommand \url  [0]{\begingroup\@sanitize@url \@url }%
\providecommand \@url [1]{\endgroup\@href {#1}{\urlprefix }}%
\providecommand \urlprefix  [0]{URL }%
\providecommand \Eprint [0]{\href }%
\providecommand \doibase [0]{http://dx.doi.org/}%
\providecommand \selectlanguage [0]{\@gobble}%
\providecommand \bibinfo  [0]{\@secondoftwo}%
\providecommand \bibfield  [0]{\@secondoftwo}%
\providecommand \translation [1]{[#1]}%
\providecommand \BibitemOpen [0]{}%
\providecommand \bibitemStop [0]{}%
\providecommand \bibitemNoStop [0]{.\EOS\space}%
\providecommand \EOS [0]{\spacefactor3000\relax}%
\providecommand \BibitemShut  [1]{\csname bibitem#1\endcsname}%
\let\auto@bib@innerbib\@empty
%</preamble>
\bibitem [{\citenamefont {Hayes}(2001)}]{Hayes2001}%
  \BibitemOpen
  \bibfield  {author} {\bibinfo {author} {\bibfnamefont {Brian}\ \bibnamefont
  {Hayes}},\ }\bibfield  {title} {\enquote {\bibinfo {title} {Randomness as a
  resource},}\ }\href {\doibase 10.1511/2001.28.3336} {\bibfield  {journal}
  {\bibinfo  {journal} {American Scientist}\ }\textbf {\bibinfo {volume}
  {89}},\ \bibinfo {pages} {300--304} (\bibinfo {year} {2001})}\BibitemShut
  {NoStop}%
\bibitem [{\citenamefont {Acin}\ and\ \citenamefont
  {Masanes}(2016)}]{Acin2016}%
  \BibitemOpen
  \bibfield  {author} {\bibinfo {author} {\bibfnamefont {A.}~\bibnamefont
  {Acin}}\ and\ \bibinfo {author} {\bibfnamefont {L.}~\bibnamefont {Masanes}},\
  }\bibfield  {title} {\enquote {\bibinfo {title} {Certified randomness in
  quantum physics},}\ }\href {\doibase doi:10.1038/nature20119} {\bibfield
  {journal} {\bibinfo  {journal} {Nature}\ }\textbf {\bibinfo {volume} {540}},\
  \bibinfo {pages} {213} (\bibinfo {year} {2016})}\BibitemShut {NoStop}%
\bibitem [{\citenamefont {Herrero-Collantes}\ and\ \citenamefont
  {Garcia-Escartin}(2017)}]{Herrero2017}%
  \BibitemOpen
  \bibfield  {author} {\bibinfo {author} {\bibfnamefont {M.}~\bibnamefont
  {Herrero-Collantes}}\ and\ \bibinfo {author} {\bibfnamefont {J.~C.}\
  \bibnamefont {Garcia-Escartin}},\ }\bibfield  {title} {\enquote {\bibinfo
  {title} {Quantum random number generators},}\ }\href {\doibase
  10.1103/RevModPhys.89.015004} {\bibfield  {journal} {\bibinfo  {journal}
  {Rev. Mod. Phys.}\ }\textbf {\bibinfo {volume} {89}},\ \bibinfo {pages}
  {015004} (\bibinfo {year} {2017})}\BibitemShut {NoStop}%
\bibitem [{\citenamefont {Bera}\ \emph {et~al.}(2017)\citenamefont {Bera},
  \citenamefont {Acin}, \citenamefont {Kus}, \citenamefont {Mitchell},\ and\
  \citenamefont {Lewenstein}}]{Bera2017}%
  \BibitemOpen
  \bibfield  {author} {\bibinfo {author} {\bibfnamefont {M.N.}\ \bibnamefont
  {Bera}}, \bibinfo {author} {\bibfnamefont {A.}~\bibnamefont {Acin}}, \bibinfo
  {author} {\bibfnamefont {M.}~\bibnamefont {Kus}}, \bibinfo {author}
  {\bibfnamefont {M.}~\bibnamefont {Mitchell}}, \ and\ \bibinfo {author}
  {\bibfnamefont {M.}~\bibnamefont {Lewenstein}},\ }\bibfield  {title}
  {\enquote {\bibinfo {title} {{Randomness in Quantum Mechanics: Philosophy,
  Physics and Technology}},}\ }\href {\doibase 10.1088/1361-6633/aa8731}
  {\bibfield  {journal} {\bibinfo  {journal} {Rep. Prog. Phys.}\ }\textbf
  {\bibinfo {volume} {80}},\ \bibinfo {pages} {124001} (\bibinfo {year}
  {2017})}\BibitemShut {NoStop}%
\bibitem [{\citenamefont {Stefanov}\ \emph {et~al.}(2000)\citenamefont
  {Stefanov}, \citenamefont {Gisin}, \citenamefont {Guinnard}, \citenamefont
  {Guinnard},\ and\ \citenamefont {Zbinden}}]{Stefanov2000}%
  \BibitemOpen
  \bibfield  {author} {\bibinfo {author} {\bibfnamefont {A.}~\bibnamefont
  {Stefanov}}, \bibinfo {author} {\bibfnamefont {N.}~\bibnamefont {Gisin}},
  \bibinfo {author} {\bibfnamefont {O.}~\bibnamefont {Guinnard}}, \bibinfo
  {author} {\bibfnamefont {L.}~\bibnamefont {Guinnard}}, \ and\ \bibinfo
  {author} {\bibfnamefont {H.}~\bibnamefont {Zbinden}},\ }\bibfield  {title}
  {\enquote {\bibinfo {title} {Optical quantum random number generator},}\
  }\href@noop {} {\bibfield  {journal} {\bibinfo  {journal} {J. Mod. Opt.}\
  }\textbf {\bibinfo {volume} {47}},\ \bibinfo {pages} {595--598} (\bibinfo
  {year} {2000})}\BibitemShut {NoStop}%
\bibitem [{\citenamefont {Bell}(1964)}]{Bell1964}%
  \BibitemOpen
  \bibfield  {author} {\bibinfo {author} {\bibfnamefont {John}\ \bibnamefont
  {Bell}},\ }\bibfield  {title} {\enquote {\bibinfo {title} {On the einstein
  podolsky rosen paradox},}\ }\href@noop {} {\bibfield  {journal} {\bibinfo
  {journal} {Physics}\ }\textbf {\bibinfo {volume} {1}},\ \bibinfo {pages}
  {195--200} (\bibinfo {year} {1964})}\BibitemShut {NoStop}%
\bibitem [{\citenamefont {Brunner}\ \emph {et~al.}(2014)\citenamefont
  {Brunner}, \citenamefont {Cavalcanti}, \citenamefont {Pironio}, \citenamefont
  {Scarani},\ and\ \citenamefont {Wehner}}]{Brunner2014}%
  \BibitemOpen
  \bibfield  {author} {\bibinfo {author} {\bibfnamefont {Nicolas}\ \bibnamefont
  {Brunner}}, \bibinfo {author} {\bibfnamefont {Daniel}\ \bibnamefont
  {Cavalcanti}}, \bibinfo {author} {\bibfnamefont {Stefano}\ \bibnamefont
  {Pironio}}, \bibinfo {author} {\bibfnamefont {Valerio}\ \bibnamefont
  {Scarani}}, \ and\ \bibinfo {author} {\bibfnamefont {Stephanie}\ \bibnamefont
  {Wehner}},\ }\bibfield  {title} {\enquote {\bibinfo {title} {Bell
  nonlocality},}\ }\href {\doibase 10.1103/RevModPhys.86.419} {\bibfield
  {journal} {\bibinfo  {journal} {Rev. Mod. Phys.}\ }\textbf {\bibinfo {volume}
  {86}},\ \bibinfo {pages} {419--478} (\bibinfo {year} {2014})}\BibitemShut
  {NoStop}%
\bibitem [{\citenamefont {Colbeck}(2009)}]{colbeckPhD2009}%
  \BibitemOpen
  \bibfield  {author} {\bibinfo {author} {\bibfnamefont {R.}~\bibnamefont
  {Colbeck}},\ }\href {https://arxiv.org/abs/0911.3814} {\enquote {\bibinfo
  {title} {Quantum and relativistic protocols for secure multi-party
  computation},}\ }\bibinfo {howpublished} {Ph.D. Thesis, University of
  Cambridge} (\bibinfo {year} {2009}),\ \bibinfo {note} {arXiv:0911.3814
  [quant-ph]}\BibitemShut {NoStop}%
\bibitem [{\citenamefont {Pironio}\ \emph {et~al.}(2010)\citenamefont
  {Pironio}, \citenamefont {Ac\'in}, \citenamefont {Massar}, \citenamefont
  {de~la Giroday}, \citenamefont {Matsukevich}, \citenamefont {Maunz},
  \citenamefont {Olmschenk}, \citenamefont {Hayes}, \citenamefont {Luo},
  \citenamefont {Manning},\ and\ \citenamefont {Monroe}}]{Pironio2010}%
  \BibitemOpen
  \bibfield  {author} {\bibinfo {author} {\bibfnamefont {S.}~\bibnamefont
  {Pironio}}, \bibinfo {author} {\bibfnamefont {A.}~\bibnamefont {Ac\'in}},
  \bibinfo {author} {\bibfnamefont {S.}~\bibnamefont {Massar}}, \bibinfo
  {author} {\bibfnamefont {A.~Boyer}\ \bibnamefont {de~la Giroday}}, \bibinfo
  {author} {\bibfnamefont {D.~N.}\ \bibnamefont {Matsukevich}}, \bibinfo
  {author} {\bibfnamefont {P.}~\bibnamefont {Maunz}}, \bibinfo {author}
  {\bibfnamefont {S.}~\bibnamefont {Olmschenk}}, \bibinfo {author}
  {\bibfnamefont {D.}~\bibnamefont {Hayes}}, \bibinfo {author} {\bibfnamefont
  {L.}~\bibnamefont {Luo}}, \bibinfo {author} {\bibfnamefont {T.~A.}\
  \bibnamefont {Manning}}, \ and\ \bibinfo {author} {\bibfnamefont
  {C.}~\bibnamefont {Monroe}},\ }\bibfield  {title} {\enquote {\bibinfo {title}
  {Random numbers certified by bell's theorem},}\ }\href {\doibase
  doi:10.1038/nature09008} {\bibfield  {journal} {\bibinfo  {journal} {Nature}\
  }\textbf {\bibinfo {volume} {464}},\ \bibinfo {pages} {1021--1024} (\bibinfo
  {year} {2010})}\BibitemShut {NoStop}%
\bibitem [{\citenamefont {Christensen}\ \emph {et~al.}(2013)\citenamefont
  {Christensen}, \citenamefont {McCusker}, \citenamefont {Altepeter},
  \citenamefont {Calkins}, \citenamefont {Gerrits}, \citenamefont {Lita},
  \citenamefont {Miller}, \citenamefont {Shalm}, \citenamefont {Zhang},
  \citenamefont {Nam}, \citenamefont {Brunner}, \citenamefont {Lim},
  \citenamefont {Gisin},\ and\ \citenamefont {Kwiat}}]{Christensen2013}%
  \BibitemOpen
  \bibfield  {author} {\bibinfo {author} {\bibfnamefont {B.~G.}\ \bibnamefont
  {Christensen}}, \bibinfo {author} {\bibfnamefont {K.~T.}\ \bibnamefont
  {McCusker}}, \bibinfo {author} {\bibfnamefont {J.~B.}\ \bibnamefont
  {Altepeter}}, \bibinfo {author} {\bibfnamefont {B.}~\bibnamefont {Calkins}},
  \bibinfo {author} {\bibfnamefont {T.}~\bibnamefont {Gerrits}}, \bibinfo
  {author} {\bibfnamefont {A.~E.}\ \bibnamefont {Lita}}, \bibinfo {author}
  {\bibfnamefont {A.}~\bibnamefont {Miller}}, \bibinfo {author} {\bibfnamefont
  {L.~K.}\ \bibnamefont {Shalm}}, \bibinfo {author} {\bibfnamefont
  {Y.}~\bibnamefont {Zhang}}, \bibinfo {author} {\bibfnamefont {S.~W.}\
  \bibnamefont {Nam}}, \bibinfo {author} {\bibfnamefont {N.}~\bibnamefont
  {Brunner}}, \bibinfo {author} {\bibfnamefont {C.~C.~W.}\ \bibnamefont {Lim}},
  \bibinfo {author} {\bibfnamefont {N.}~\bibnamefont {Gisin}}, \ and\ \bibinfo
  {author} {\bibfnamefont {P.~G.}\ \bibnamefont {Kwiat}},\ }\bibfield  {title}
  {\enquote {\bibinfo {title} {Detection-loophole-free test of quantum
  nonlocality, and applications},}\ }\href {\doibase
  10.1103/PhysRevLett.111.130406} {\bibfield  {journal} {\bibinfo  {journal}
  {Phys. Rev. Lett.}\ }\textbf {\bibinfo {volume} {111}},\ \bibinfo {pages}
  {130406} (\bibinfo {year} {2013})}\BibitemShut {NoStop}%
\bibitem [{\citenamefont {Bierhorst}\ \emph {et~al.}(2018)\citenamefont
  {Bierhorst}, \citenamefont {Knill}, \citenamefont {Glancy}, \citenamefont
  {Zhang}, \citenamefont {Mink}, \citenamefont {Jordan}, \citenamefont
  {Rommal}, \citenamefont {Liu}, \citenamefont {Christensen}, \citenamefont
  {Nam}, \citenamefont {Stevens},\ and\ \citenamefont {Shalm}}]{Bierhorst2018}%
  \BibitemOpen
  \bibfield  {author} {\bibinfo {author} {\bibfnamefont {P.}~\bibnamefont
  {Bierhorst}}, \bibinfo {author} {\bibfnamefont {E.}~\bibnamefont {Knill}},
  \bibinfo {author} {\bibfnamefont {S.}~\bibnamefont {Glancy}}, \bibinfo
  {author} {\bibfnamefont {Y.}~\bibnamefont {Zhang}}, \bibinfo {author}
  {\bibfnamefont {A.}~\bibnamefont {Mink}}, \bibinfo {author} {\bibfnamefont
  {S.}~\bibnamefont {Jordan}}, \bibinfo {author} {\bibfnamefont
  {A.}~\bibnamefont {Rommal}}, \bibinfo {author} {\bibfnamefont {Y.-K.}\
  \bibnamefont {Liu}}, \bibinfo {author} {\bibfnamefont {B.}~\bibnamefont
  {Christensen}}, \bibinfo {author} {\bibfnamefont {S.~W.}\ \bibnamefont
  {Nam}}, \bibinfo {author} {\bibfnamefont {M.~J.}\ \bibnamefont {Stevens}}, \
  and\ \bibinfo {author} {\bibfnamefont {L.~K.}\ \bibnamefont {Shalm}},\
  }\bibfield  {title} {\enquote {\bibinfo {title} {Experimentally {Generated}
  {Randomness} {Certified} by the {Impossibility} of {Superluminal}
  {Signals}},}\ }\href {\doibase 10.1038/s41586-018-0019-0} {\bibfield
  {journal} {\bibinfo  {journal} {Nature}\ }\textbf {\bibinfo {volume} {556}},\
  \bibinfo {pages} {223--226} (\bibinfo {year} {2018})}\BibitemShut {NoStop}%
\bibitem [{\citenamefont {Liu}\ \emph {et~al.}(2018)\citenamefont {Liu},
  \citenamefont {Zhao}, \citenamefont {Li}, \citenamefont {Guan}, \citenamefont
  {Zhang}, \citenamefont {Bai}, \citenamefont {Zhang}, \citenamefont {Liu},
  \citenamefont {Wu}, \citenamefont {Yuan}, \citenamefont {Li}, \citenamefont
  {Munro}, \citenamefont {Wang}, \citenamefont {You}, \citenamefont {Zhang},
  \citenamefont {Ma}, \citenamefont {Fan}, \citenamefont {Zhang},\ and\
  \citenamefont {Pan}}]{Liu2018}%
  \BibitemOpen
  \bibfield  {author} {\bibinfo {author} {\bibfnamefont {Y.}~\bibnamefont
  {Liu}}, \bibinfo {author} {\bibfnamefont {Q.}~\bibnamefont {Zhao}}, \bibinfo
  {author} {\bibfnamefont {M.-H.}\ \bibnamefont {Li}}, \bibinfo {author}
  {\bibfnamefont {J.-Y.}\ \bibnamefont {Guan}}, \bibinfo {author}
  {\bibfnamefont {Y.}~\bibnamefont {Zhang}}, \bibinfo {author} {\bibfnamefont
  {B.}~\bibnamefont {Bai}}, \bibinfo {author} {\bibfnamefont {W.}~\bibnamefont
  {Zhang}}, \bibinfo {author} {\bibfnamefont {W.-Z.}\ \bibnamefont {Liu}},
  \bibinfo {author} {\bibfnamefont {C.}~\bibnamefont {Wu}}, \bibinfo {author}
  {\bibfnamefont {X.}~\bibnamefont {Yuan}}, \bibinfo {author} {\bibfnamefont
  {H.}~\bibnamefont {Li}}, \bibinfo {author} {\bibfnamefont {W.~J.}\
  \bibnamefont {Munro}}, \bibinfo {author} {\bibfnamefont {Z.}~\bibnamefont
  {Wang}}, \bibinfo {author} {\bibfnamefont {L.}~\bibnamefont {You}}, \bibinfo
  {author} {\bibfnamefont {J.}~\bibnamefont {Zhang}}, \bibinfo {author}
  {\bibfnamefont {X.}~\bibnamefont {Ma}}, \bibinfo {author} {\bibfnamefont
  {J.}~\bibnamefont {Fan}}, \bibinfo {author} {\bibfnamefont {Q.}~\bibnamefont
  {Zhang}}, \ and\ \bibinfo {author} {\bibfnamefont {J.-W.}\ \bibnamefont
  {Pan}},\ }\bibfield  {title} {\enquote {\bibinfo {title} {Device-independent
  quantum random-number generation},}\ }\href {\doibase
  10.1038/s41586-018-0559-3} {\bibfield  {journal} {\bibinfo  {journal}
  {Nature}\ }\textbf {\bibinfo {volume} {562}},\ \bibinfo {pages} {548}
  (\bibinfo {year} {2018})}\BibitemShut {NoStop}%
\bibitem [{\citenamefont {Shalm}\ \emph {et~al.}(2021)\citenamefont {Shalm},
  \citenamefont {Zhang}, \citenamefont {Bienfang}, \citenamefont {Schlager},
  \citenamefont {Stevens}, \citenamefont {Mazurek}, \citenamefont {Abellán},
  \citenamefont {Amaya}, \citenamefont {Mitchell}, \citenamefont {Alhejji},
  \citenamefont {Fu}, \citenamefont {Ornstein}, \citenamefont {Mirin},
  \citenamefont {Nam},\ and\ \citenamefont {Knill}}]{Shalm2021}%
  \BibitemOpen
  \bibfield  {author} {\bibinfo {author} {\bibfnamefont {Lynden~K.}\
  \bibnamefont {Shalm}}, \bibinfo {author} {\bibfnamefont {Yanbao}\
  \bibnamefont {Zhang}}, \bibinfo {author} {\bibfnamefont {Joshua~C.}\
  \bibnamefont {Bienfang}}, \bibinfo {author} {\bibfnamefont {Collin}\
  \bibnamefont {Schlager}}, \bibinfo {author} {\bibfnamefont {Martin~J.}\
  \bibnamefont {Stevens}}, \bibinfo {author} {\bibfnamefont {Michael~D.}\
  \bibnamefont {Mazurek}}, \bibinfo {author} {\bibfnamefont {Carlos}\
  \bibnamefont {Abellán}}, \bibinfo {author} {\bibfnamefont {Waldimar}\
  \bibnamefont {Amaya}}, \bibinfo {author} {\bibfnamefont {Morgan~W.}\
  \bibnamefont {Mitchell}}, \bibinfo {author} {\bibfnamefont {Mohammad~A.}\
  \bibnamefont {Alhejji}}, \bibinfo {author} {\bibfnamefont {Honghao}\
  \bibnamefont {Fu}}, \bibinfo {author} {\bibfnamefont {Joel}\ \bibnamefont
  {Ornstein}}, \bibinfo {author} {\bibfnamefont {Richard~P.}\ \bibnamefont
  {Mirin}}, \bibinfo {author} {\bibfnamefont {Sae~Woo}\ \bibnamefont {Nam}}, \
  and\ \bibinfo {author} {\bibfnamefont {Emanuel}\ \bibnamefont {Knill}},\
  }\bibfield  {title} {\enquote {\bibinfo {title} {Device-independent
  randomness expansion with entangled photons},}\ }\href {\doibase
  10.1038/s41567-020-01153-4} {\bibfield  {journal} {\bibinfo  {journal}
  {Nature Physics}\ }\textbf {\bibinfo {volume} {17}},\ \bibinfo {pages}
  {452--456} (\bibinfo {year} {2021})}\BibitemShut {NoStop}%
\bibitem [{\citenamefont {Liu}\ \emph {et~al.}(2021)\citenamefont {Liu},
  \citenamefont {Li}, \citenamefont {Ragy}, \citenamefont {Zhao}, \citenamefont
  {Bai}, \citenamefont {Liu}, \citenamefont {Brown}, \citenamefont {Zhang},
  \citenamefont {Colbeck}, \citenamefont {Fan}, \citenamefont {Zhang},\ and\
  \citenamefont {Pan}}]{Liu2021}%
  \BibitemOpen
  \bibfield  {author} {\bibinfo {author} {\bibfnamefont {Wen-Zhao}\
  \bibnamefont {Liu}}, \bibinfo {author} {\bibfnamefont {Ming-Han}\
  \bibnamefont {Li}}, \bibinfo {author} {\bibfnamefont {Sammy}\ \bibnamefont
  {Ragy}}, \bibinfo {author} {\bibfnamefont {Si-Ran}\ \bibnamefont {Zhao}},
  \bibinfo {author} {\bibfnamefont {Bing}\ \bibnamefont {Bai}}, \bibinfo
  {author} {\bibfnamefont {Yang}\ \bibnamefont {Liu}}, \bibinfo {author}
  {\bibfnamefont {Peter~J.}\ \bibnamefont {Brown}}, \bibinfo {author}
  {\bibfnamefont {Jun}\ \bibnamefont {Zhang}}, \bibinfo {author} {\bibfnamefont
  {Roger}\ \bibnamefont {Colbeck}}, \bibinfo {author} {\bibfnamefont {Jingyun}\
  \bibnamefont {Fan}}, \bibinfo {author} {\bibfnamefont {Qiang}\ \bibnamefont
  {Zhang}}, \ and\ \bibinfo {author} {\bibfnamefont {Jian-Wei}\ \bibnamefont
  {Pan}},\ }\bibfield  {title} {\enquote {\bibinfo {title} {Device-independent
  randomness expansion against quantum side information},}\ }\href {\doibase
  10.1038/s41567-020-01147-2} {\bibfield  {journal} {\bibinfo  {journal}
  {Nature Physics}\ }\textbf {\bibinfo {volume} {17}},\ \bibinfo {pages}
  {448--451} (\bibinfo {year} {2021})}\BibitemShut {NoStop}%
\bibitem [{\citenamefont {Li}\ \emph {et~al.}(2011)\citenamefont {Li},
  \citenamefont {Yin}, \citenamefont {Wu}, \citenamefont {Zou}, \citenamefont
  {Wang}, \citenamefont {Chen}, \citenamefont {Guo},\ and\ \citenamefont
  {Han}}]{Li2011}%
  \BibitemOpen
  \bibfield  {author} {\bibinfo {author} {\bibfnamefont {H.-W.}\ \bibnamefont
  {Li}}, \bibinfo {author} {\bibfnamefont {Z.-Q.}\ \bibnamefont {Yin}},
  \bibinfo {author} {\bibfnamefont {Y.-C.}\ \bibnamefont {Wu}}, \bibinfo
  {author} {\bibfnamefont {X.-B.}\ \bibnamefont {Zou}}, \bibinfo {author}
  {\bibfnamefont {S.}~\bibnamefont {Wang}}, \bibinfo {author} {\bibfnamefont
  {W.}~\bibnamefont {Chen}}, \bibinfo {author} {\bibfnamefont {G.-C.}\
  \bibnamefont {Guo}}, \ and\ \bibinfo {author} {\bibfnamefont {Z.-F.}\
  \bibnamefont {Han}},\ }\bibfield  {title} {\enquote {\bibinfo {title}
  {Semi-device-independent random-number expansion without entanglement},}\
  }\href {\doibase 10.1103/PhysRevA.84.034301} {\bibfield  {journal} {\bibinfo
  {journal} {Phys. Rev. A}\ }\textbf {\bibinfo {volume} {84}},\ \bibinfo
  {pages} {034301} (\bibinfo {year} {2011})}\BibitemShut {NoStop}%
\bibitem [{\citenamefont {Vallone}\ \emph {et~al.}(2014)\citenamefont
  {Vallone}, \citenamefont {Marangon}, \citenamefont {Tomasin},\ and\
  \citenamefont {Villoresi}}]{Vallone2014}%
  \BibitemOpen
  \bibfield  {author} {\bibinfo {author} {\bibfnamefont {G.}~\bibnamefont
  {Vallone}}, \bibinfo {author} {\bibfnamefont {D.~G.}\ \bibnamefont
  {Marangon}}, \bibinfo {author} {\bibfnamefont {M.}~\bibnamefont {Tomasin}}, \
  and\ \bibinfo {author} {\bibfnamefont {P.}~\bibnamefont {Villoresi}},\
  }\bibfield  {title} {\enquote {\bibinfo {title} {{Quantum randomness
  certified by the uncertainty principle}},}\ }\href {\doibase
  10.1103/PhysRevA.90.052327} {\bibfield  {journal} {\bibinfo  {journal} {Phys.
  Rev. A}\ }\textbf {\bibinfo {volume} {90}},\ \bibinfo {pages} {052327}
  (\bibinfo {year} {2014})}\BibitemShut {NoStop}%
\bibitem [{\citenamefont {Lunghi}\ \emph {et~al.}(2015)\citenamefont {Lunghi},
  \citenamefont {Brask}, \citenamefont {Lim}, \citenamefont {Lavigne},
  \citenamefont {Bowles}, \citenamefont {Martin}, \citenamefont {Zbinden},\
  and\ \citenamefont {Brunner}}]{Lunghi2015}%
  \BibitemOpen
  \bibfield  {author} {\bibinfo {author} {\bibfnamefont {T.}~\bibnamefont
  {Lunghi}}, \bibinfo {author} {\bibfnamefont {J.~B.}\ \bibnamefont {Brask}},
  \bibinfo {author} {\bibfnamefont {C.~C.~W.}\ \bibnamefont {Lim}}, \bibinfo
  {author} {\bibfnamefont {Q.}~\bibnamefont {Lavigne}}, \bibinfo {author}
  {\bibfnamefont {J.}~\bibnamefont {Bowles}}, \bibinfo {author} {\bibfnamefont
  {A.}~\bibnamefont {Martin}}, \bibinfo {author} {\bibfnamefont
  {H.}~\bibnamefont {Zbinden}}, \ and\ \bibinfo {author} {\bibfnamefont
  {N.}~\bibnamefont {Brunner}},\ }\bibfield  {title} {\enquote {\bibinfo
  {title} {Self-testing quantum random number generator},}\ }\href {\doibase
  10.1103/PhysRevLett.114.150501} {\bibfield  {journal} {\bibinfo  {journal}
  {Phys. Rev. Lett.}\ }\textbf {\bibinfo {volume} {114}},\ \bibinfo {pages}
  {150501} (\bibinfo {year} {2015})}\BibitemShut {NoStop}%
\bibitem [{\citenamefont {Mironowicz}\ \emph {et~al.}(2021)\citenamefont
  {Mironowicz}, \citenamefont {Cañas}, \citenamefont {Cariñe}, \citenamefont
  {Gómez}, \citenamefont {Barra}, \citenamefont {Cabello}, \citenamefont
  {Xavier}, \citenamefont {Lima},\ and\ \citenamefont
  {Pawłowski}}]{Mironowicz2021}%
  \BibitemOpen
  \bibfield  {author} {\bibinfo {author} {\bibfnamefont {P.}~\bibnamefont
  {Mironowicz}}, \bibinfo {author} {\bibfnamefont {G.}~\bibnamefont {Cañas}},
  \bibinfo {author} {\bibfnamefont {J.}~\bibnamefont {Cariñe}}, \bibinfo
  {author} {\bibfnamefont {E.~S.}\ \bibnamefont {Gómez}}, \bibinfo {author}
  {\bibfnamefont {J.~F.}\ \bibnamefont {Barra}}, \bibinfo {author}
  {\bibfnamefont {A.}~\bibnamefont {Cabello}}, \bibinfo {author} {\bibfnamefont
  {G.~B.}\ \bibnamefont {Xavier}}, \bibinfo {author} {\bibfnamefont
  {G.}~\bibnamefont {Lima}}, \ and\ \bibinfo {author} {\bibfnamefont
  {M.}~\bibnamefont {Pawłowski}},\ }\bibfield  {title} {\enquote {\bibinfo
  {title} {Quantum randomness protected against detection loophole attacks},}\
  }\href {\doibase 10.1007/s11128-020-02948-3} {\bibfield  {journal} {\bibinfo
  {journal} {Quantum Information Processing}\ }\textbf {\bibinfo {volume}
  {20}},\ \bibinfo {pages} {39} (\bibinfo {year} {2021})}\BibitemShut {NoStop}%
\bibitem [{\citenamefont {Cao}\ \emph {et~al.}(2015)\citenamefont {Cao},
  \citenamefont {Zhou},\ and\ \citenamefont {Ma}}]{Cao2015}%
  \BibitemOpen
  \bibfield  {author} {\bibinfo {author} {\bibfnamefont {Z.}~\bibnamefont
  {Cao}}, \bibinfo {author} {\bibfnamefont {H.}~\bibnamefont {Zhou}}, \ and\
  \bibinfo {author} {\bibfnamefont {X.}~\bibnamefont {Ma}},\ }\bibfield
  {title} {\enquote {\bibinfo {title} {{Loss-tolerant
  measurement-device-independent quantum random number generation}},}\ }\href
  {\doibase 10.1088/1367-2630/17/12/125011} {\bibfield  {journal} {\bibinfo
  {journal} {New J. Phys.}\ }\textbf {\bibinfo {volume} {17}},\ \bibinfo
  {pages} {125011} (\bibinfo {year} {2015})}\BibitemShut {NoStop}%
\bibitem [{\citenamefont {Marangon}\ \emph {et~al.}(2017)\citenamefont
  {Marangon}, \citenamefont {Vallone},\ and\ \citenamefont
  {Villoresi}}]{Marangon2017}%
  \BibitemOpen
  \bibfield  {author} {\bibinfo {author} {\bibfnamefont {D.~G.}\ \bibnamefont
  {Marangon}}, \bibinfo {author} {\bibfnamefont {G.}~\bibnamefont {Vallone}}, \
  and\ \bibinfo {author} {\bibfnamefont {P.}~\bibnamefont {Villoresi}},\
  }\bibfield  {title} {\enquote {\bibinfo {title} {Source-device-independent
  ultrafast quantum random number generation},}\ }\href {\doibase
  10.1103/PhysRevLett.118.060503} {\bibfield  {journal} {\bibinfo  {journal}
  {Phys. Rev. Lett.}\ }\textbf {\bibinfo {volume} {118}},\ \bibinfo {pages}
  {060503} (\bibinfo {year} {2017})}\BibitemShut {NoStop}%
\bibitem [{\citenamefont {Cao}\ \emph {et~al.}(2016)\citenamefont {Cao},
  \citenamefont {Zhou}, \citenamefont {Yuan},\ and\ \citenamefont
  {Ma}}]{Cao2016}%
  \BibitemOpen
  \bibfield  {author} {\bibinfo {author} {\bibfnamefont {Z.}~\bibnamefont
  {Cao}}, \bibinfo {author} {\bibfnamefont {H.}~\bibnamefont {Zhou}}, \bibinfo
  {author} {\bibfnamefont {X.}~\bibnamefont {Yuan}}, \ and\ \bibinfo {author}
  {\bibfnamefont {X.}~\bibnamefont {Ma}},\ }\bibfield  {title} {\enquote
  {\bibinfo {title} {{Source-Independent Quantum Random Number Generation}},}\
  }\href {\doibase 10.1103/PhysRevX.6.011020} {\bibfield  {journal} {\bibinfo
  {journal} {Phys. Rev. X}\ }\textbf {\bibinfo {volume} {6}},\ \bibinfo {pages}
  {011020} (\bibinfo {year} {2016})}\BibitemShut {NoStop}%
\bibitem [{\citenamefont {Xu}\ \emph {et~al.}(2016)\citenamefont {Xu},
  \citenamefont {Shapiro},\ and\ \citenamefont {Wong}}]{Xu2016}%
  \BibitemOpen
  \bibfield  {author} {\bibinfo {author} {\bibfnamefont {F}~\bibnamefont {Xu}},
  \bibinfo {author} {\bibfnamefont {J.~H.}\ \bibnamefont {Shapiro}}, \ and\
  \bibinfo {author} {\bibfnamefont {F.~N.~C.}\ \bibnamefont {Wong}},\
  }\bibfield  {title} {\enquote {\bibinfo {title} {Experimental fast quantum
  random number generation using high-dimensional entanglement with entropy
  monitoring},}\ }\href {\doibase 10.1364/OPTICA.3.001266} {\bibfield
  {journal} {\bibinfo  {journal} {Optica}\ }\textbf {\bibinfo {volume} {3}},\
  \bibinfo {pages} {1266--1269} (\bibinfo {year} {2016})}\BibitemShut {NoStop}%
\bibitem [{\citenamefont {Brask}\ \emph {et~al.}(2017)\citenamefont {Brask},
  \citenamefont {Martin}, \citenamefont {Esposito}, \citenamefont {Houlmann},
  \citenamefont {Bowles}, \citenamefont {Zbinden},\ and\ \citenamefont
  {Brunner}}]{Brask2017a}%
  \BibitemOpen
  \bibfield  {author} {\bibinfo {author} {\bibfnamefont {J.~B.}\ \bibnamefont
  {Brask}}, \bibinfo {author} {\bibfnamefont {A.}~\bibnamefont {Martin}},
  \bibinfo {author} {\bibfnamefont {W.}~\bibnamefont {Esposito}}, \bibinfo
  {author} {\bibfnamefont {R.}~\bibnamefont {Houlmann}}, \bibinfo {author}
  {\bibfnamefont {J.}~\bibnamefont {Bowles}}, \bibinfo {author} {\bibfnamefont
  {H.}~\bibnamefont {Zbinden}}, \ and\ \bibinfo {author} {\bibfnamefont
  {N.}~\bibnamefont {Brunner}},\ }\bibfield  {title} {\enquote {\bibinfo
  {title} {{Megahertz-Rate Semi-Device-Independent Quantum Random Number
  Generators Based on Unambiguous State Discrimination}},}\ }\href {\doibase
  10.1103/PhysRevApplied.7.054018} {\bibfield  {journal} {\bibinfo  {journal}
  {Phys. Rev. Appl.}\ }\textbf {\bibinfo {volume} {7}},\ \bibinfo {pages}
  {054018} (\bibinfo {year} {2017})}\BibitemShut {NoStop}%
\bibitem [{\citenamefont {Gehring}\ \emph {et~al.}(2021)\citenamefont
  {Gehring}, \citenamefont {Lupo}, \citenamefont {Kordts}, \citenamefont
  {Solar~Nikolic}, \citenamefont {Jain}, \citenamefont {Rydberg}, \citenamefont
  {Pedersen}, \citenamefont {Pirandola},\ and\ \citenamefont
  {Andersen}}]{Gehring2021}%
  \BibitemOpen
  \bibfield  {author} {\bibinfo {author} {\bibfnamefont {Tobias}\ \bibnamefont
  {Gehring}}, \bibinfo {author} {\bibfnamefont {Cosmo}\ \bibnamefont {Lupo}},
  \bibinfo {author} {\bibfnamefont {Arne}\ \bibnamefont {Kordts}}, \bibinfo
  {author} {\bibfnamefont {Dino}\ \bibnamefont {Solar~Nikolic}}, \bibinfo
  {author} {\bibfnamefont {Nitin}\ \bibnamefont {Jain}}, \bibinfo {author}
  {\bibfnamefont {Tobias}\ \bibnamefont {Rydberg}}, \bibinfo {author}
  {\bibfnamefont {Thomas~B.}\ \bibnamefont {Pedersen}}, \bibinfo {author}
  {\bibfnamefont {Stefano}\ \bibnamefont {Pirandola}}, \ and\ \bibinfo {author}
  {\bibfnamefont {Ulrik~L.}\ \bibnamefont {Andersen}},\ }\bibfield  {title}
  {\enquote {\bibinfo {title} {Homodyne-based quantum random number generator
  at 2.9 gbps secure against quantum side-information},}\ }\href {\doibase
  10.1038/s41467-020-20813-w} {\bibfield  {journal} {\bibinfo  {journal}
  {Nature Communications}\ }\textbf {\bibinfo {volume} {12}},\ \bibinfo {pages}
  {605} (\bibinfo {year} {2021})}\BibitemShut {NoStop}%
\bibitem [{\citenamefont {Michel}\ \emph {et~al.}(2019)\citenamefont {Michel},
  \citenamefont {Haw}, \citenamefont {Marangon}, \citenamefont {Thearle},
  \citenamefont {Vallone}, \citenamefont {Villoresi}, \citenamefont {Lam},\
  and\ \citenamefont {Assad}}]{Michel2019}%
  \BibitemOpen
  \bibfield  {author} {\bibinfo {author} {\bibfnamefont {T.}~\bibnamefont
  {Michel}}, \bibinfo {author} {\bibfnamefont {J.}~\bibnamefont {Haw}},
  \bibinfo {author} {\bibfnamefont {D.}~\bibnamefont {Marangon}}, \bibinfo
  {author} {\bibfnamefont {O.}~\bibnamefont {Thearle}}, \bibinfo {author}
  {\bibfnamefont {G.}~\bibnamefont {Vallone}}, \bibinfo {author} {\bibfnamefont
  {P.}~\bibnamefont {Villoresi}}, \bibinfo {author} {\bibfnamefont {P.K.}\
  \bibnamefont {Lam}}, \ and\ \bibinfo {author} {\bibfnamefont {S.M.}\
  \bibnamefont {Assad}},\ }\bibfield  {title} {\enquote {\bibinfo {title}
  {Real-time source independent quantum random number generator with squeezed
  states},}\ }\href@noop {} {\bibfield  {journal} {\bibinfo  {journal}
  {arXiv:1903.01071}\ } (\bibinfo {year} {2019})}\BibitemShut {NoStop}%
\bibitem [{\citenamefont {Rusca}\ \emph {et~al.}(2019)\citenamefont {Rusca},
  \citenamefont {van Himbeeck}, \citenamefont {Martin}, \citenamefont {Brask},
  \citenamefont {Shi}, \citenamefont {Pironio}, \citenamefont {Brunner},\ and\
  \citenamefont {Zbinden}}]{Rusca2019}%
  \BibitemOpen
  \bibfield  {author} {\bibinfo {author} {\bibfnamefont {D.}~\bibnamefont
  {Rusca}}, \bibinfo {author} {\bibfnamefont {T.}~\bibnamefont {van Himbeeck}},
  \bibinfo {author} {\bibfnamefont {A.}~\bibnamefont {Martin}}, \bibinfo
  {author} {\bibfnamefont {J.~B.}\ \bibnamefont {Brask}}, \bibinfo {author}
  {\bibfnamefont {W.}~\bibnamefont {Shi}}, \bibinfo {author} {\bibfnamefont
  {S.}~\bibnamefont {Pironio}}, \bibinfo {author} {\bibfnamefont
  {N.}~\bibnamefont {Brunner}}, \ and\ \bibinfo {author} {\bibfnamefont
  {H.}~\bibnamefont {Zbinden}},\ }\bibfield  {title} {\enquote {\bibinfo
  {title} {Self-testing quantum random-number generator based on an energy
  bound},}\ }\href {\doibase 10.1103/PhysRevA.100.062338} {\bibfield  {journal}
  {\bibinfo  {journal} {Phys. Rev. A}\ }\textbf {\bibinfo {volume} {100}},\
  \bibinfo {pages} {062338} (\bibinfo {year} {2019})}\BibitemShut {NoStop}%
\bibitem [{\citenamefont {Drahi}\ \emph {et~al.}(2020)\citenamefont {Drahi},
  \citenamefont {Walk}, \citenamefont {Hoban}, \citenamefont {Fedorov},
  \citenamefont {Shakhovoy}, \citenamefont {Feimov}, \citenamefont {Kurochkin},
  \citenamefont {Kolthammer}, \citenamefont {Nunn}, \citenamefont {Barrett},\
  and\ \citenamefont {Walmsley}}]{Drahi2020}%
  \BibitemOpen
  \bibfield  {author} {\bibinfo {author} {\bibfnamefont {D.}~\bibnamefont
  {Drahi}}, \bibinfo {author} {\bibfnamefont {N.}~\bibnamefont {Walk}},
  \bibinfo {author} {\bibfnamefont {M.~J.}\ \bibnamefont {Hoban}}, \bibinfo
  {author} {\bibfnamefont {A.~K.}\ \bibnamefont {Fedorov}}, \bibinfo {author}
  {\bibfnamefont {R.}~\bibnamefont {Shakhovoy}}, \bibinfo {author}
  {\bibfnamefont {A.}~\bibnamefont {Feimov}}, \bibinfo {author} {\bibfnamefont
  {Y.}~\bibnamefont {Kurochkin}}, \bibinfo {author} {\bibfnamefont {W.~S.}\
  \bibnamefont {Kolthammer}}, \bibinfo {author} {\bibfnamefont
  {J.}~\bibnamefont {Nunn}}, \bibinfo {author} {\bibfnamefont {J.}~\bibnamefont
  {Barrett}}, \ and\ \bibinfo {author} {\bibfnamefont {I.~A.}\ \bibnamefont
  {Walmsley}},\ }\bibfield  {title} {\enquote {\bibinfo {title} {Certified
  quantum random numbers from untrusted light},}\ }\href {\doibase
  10.1103/PhysRevX.10.041048} {\bibfield  {journal} {\bibinfo  {journal} {Phys.
  Rev. X}\ }\textbf {\bibinfo {volume} {10}},\ \bibinfo {pages} {041048}
  (\bibinfo {year} {2020})}\BibitemShut {NoStop}%
\bibitem [{\citenamefont {Rusca}\ \emph {et~al.}(2020)\citenamefont {Rusca},
  \citenamefont {Tebyanian}, \citenamefont {Martin},\ and\ \citenamefont
  {Zbinden}}]{Rusca2020}%
  \BibitemOpen
  \bibfield  {author} {\bibinfo {author} {\bibfnamefont {D.}~\bibnamefont
  {Rusca}}, \bibinfo {author} {\bibfnamefont {H.}~\bibnamefont {Tebyanian}},
  \bibinfo {author} {\bibfnamefont {A.}~\bibnamefont {Martin}}, \ and\ \bibinfo
  {author} {\bibfnamefont {H.}~\bibnamefont {Zbinden}},\ }\bibfield  {title}
  {\enquote {\bibinfo {title} {Fast self-testing quantum random number
  generator based on homodyne detection},}\ }\href {\doibase 10.1063/5.0011479}
  {\bibfield  {journal} {\bibinfo  {journal} {Applied Physics Letters}\
  }\textbf {\bibinfo {volume} {116}},\ \bibinfo {pages} {264004} (\bibinfo
  {year} {2020})}\BibitemShut {NoStop}%
\bibitem [{\citenamefont {Avesani}\ \emph {et~al.}(2021)\citenamefont
  {Avesani}, \citenamefont {Tebyanian}, \citenamefont {Villoresi},\ and\
  \citenamefont {Vallone}}]{Avesani2021}%
  \BibitemOpen
  \bibfield  {author} {\bibinfo {author} {\bibfnamefont {M.}~\bibnamefont
  {Avesani}}, \bibinfo {author} {\bibfnamefont {H.}~\bibnamefont {Tebyanian}},
  \bibinfo {author} {\bibfnamefont {P.}~\bibnamefont {Villoresi}}, \ and\
  \bibinfo {author} {\bibfnamefont {G.}~\bibnamefont {Vallone}},\ }\bibfield
  {title} {\enquote {\bibinfo {title} {Semi-device-independent heterodyne-based
  quantum random-number generator},}\ }\href {\doibase
  10.1103/PhysRevApplied.15.034034} {\bibfield  {journal} {\bibinfo  {journal}
  {Phys. Rev. Applied}\ }\textbf {\bibinfo {volume} {15}},\ \bibinfo {pages}
  {034034} (\bibinfo {year} {2021})}\BibitemShut {NoStop}%
\bibitem [{\citenamefont {Einstein}\ \emph {et~al.}(1935)\citenamefont
  {Einstein}, \citenamefont {Podolsky},\ and\ \citenamefont
  {Rosen}}]{Einstein1935}%
  \BibitemOpen
  \bibfield  {author} {\bibinfo {author} {\bibfnamefont {A.}~\bibnamefont
  {Einstein}}, \bibinfo {author} {\bibfnamefont {B.}~\bibnamefont {Podolsky}},
  \ and\ \bibinfo {author} {\bibfnamefont {N.}~\bibnamefont {Rosen}},\
  }\bibfield  {title} {\enquote {\bibinfo {title} {Can quantum-mechanical
  description of physical reality be considered complete?}}\ }\href {\doibase
  10.1103/PhysRev.47.777} {\bibfield  {journal} {\bibinfo  {journal} {Phys.
  Rev.}\ }\textbf {\bibinfo {volume} {47}},\ \bibinfo {pages} {777--780}
  (\bibinfo {year} {1935})}\BibitemShut {NoStop}%
\bibitem [{\citenamefont {Reid}(1989)}]{Reid1989}%
  \BibitemOpen
  \bibfield  {author} {\bibinfo {author} {\bibfnamefont {M.~D.}\ \bibnamefont
  {Reid}},\ }\bibfield  {title} {\enquote {\bibinfo {title} {Demonstration of
  the einstein-podolsky-rosen paradox using nondegenerate parametric
  amplification},}\ }\href {\doibase 10.1103/PhysRevA.40.913} {\bibfield
  {journal} {\bibinfo  {journal} {Phys. Rev. A}\ }\textbf {\bibinfo {volume}
  {40}},\ \bibinfo {pages} {913--923} (\bibinfo {year} {1989})}\BibitemShut
  {NoStop}%
\bibitem [{\citenamefont {Wiseman}\ \emph {et~al.}(2007)\citenamefont
  {Wiseman}, \citenamefont {Jones},\ and\ \citenamefont
  {Doherty}}]{Wiseman2007}%
  \BibitemOpen
  \bibfield  {author} {\bibinfo {author} {\bibfnamefont {H.~M.}\ \bibnamefont
  {Wiseman}}, \bibinfo {author} {\bibfnamefont {S.~J.}\ \bibnamefont {Jones}},
  \ and\ \bibinfo {author} {\bibfnamefont {A.~C.}\ \bibnamefont {Doherty}},\
  }\bibfield  {title} {\enquote {\bibinfo {title} {Steering, entanglement,
  nonlocality, and the einstein-podolsky-rosen paradox},}\ }\href {\doibase
  10.1103/PhysRevLett.98.140402} {\bibfield  {journal} {\bibinfo  {journal}
  {Phys. Rev. Lett.}\ }\textbf {\bibinfo {volume} {98}},\ \bibinfo {pages}
  {140402} (\bibinfo {year} {2007})}\BibitemShut {NoStop}%
\bibitem [{\citenamefont {Reid}(2000)}]{Reid2000}%
  \BibitemOpen
  \bibfield  {author} {\bibinfo {author} {\bibfnamefont {M.~D.}\ \bibnamefont
  {Reid}},\ }\bibfield  {title} {\enquote {\bibinfo {title} {Quantum
  cryptography with a predetermined key, using continuous-variable
  einstein-podolsky-rosen correlations},}\ }\href {\doibase
  10.1103/PhysRevA.62.062308} {\bibfield  {journal} {\bibinfo  {journal} {Phys.
  Rev. A}\ }\textbf {\bibinfo {volume} {62}},\ \bibinfo {pages} {062308}
  (\bibinfo {year} {2000})}\BibitemShut {NoStop}%
\bibitem [{\citenamefont {Branciard}\ \emph {et~al.}(2012)\citenamefont
  {Branciard}, \citenamefont {Cavalcanti}, \citenamefont {Walborn},
  \citenamefont {Scarani},\ and\ \citenamefont {Wiseman}}]{Branciard2012}%
  \BibitemOpen
  \bibfield  {author} {\bibinfo {author} {\bibfnamefont {C.}~\bibnamefont
  {Branciard}}, \bibinfo {author} {\bibfnamefont {E.~G.}\ \bibnamefont
  {Cavalcanti}}, \bibinfo {author} {\bibfnamefont {S.~P.}\ \bibnamefont
  {Walborn}}, \bibinfo {author} {\bibfnamefont {V.}~\bibnamefont {Scarani}}, \
  and\ \bibinfo {author} {\bibfnamefont {H.~M.}\ \bibnamefont {Wiseman}},\
  }\bibfield  {title} {\enquote {\bibinfo {title} {One-sided device-independent
  quantum key distribution: Security, feasibility, and the connection with
  steering},}\ }\href {\doibase 10.1103/PhysRevA.85.010301} {\bibfield
  {journal} {\bibinfo  {journal} {Phys. Rev. A}\ }\textbf {\bibinfo {volume}
  {85}},\ \bibinfo {pages} {010301(R)} (\bibinfo {year} {2012})}\BibitemShut
  {NoStop}%
\bibitem [{\citenamefont {Law}\ \emph {et~al.}(2014)\citenamefont {Law},
  \citenamefont {Thinh}, \citenamefont {Bancal},\ and\ \citenamefont
  {Scarani}}]{Law2014}%
  \BibitemOpen
  \bibfield  {author} {\bibinfo {author} {\bibfnamefont {Yun~Zhi}\ \bibnamefont
  {Law}}, \bibinfo {author} {\bibfnamefont {Le~Phuc}\ \bibnamefont {Thinh}},
  \bibinfo {author} {\bibfnamefont {Jean-Daniel}\ \bibnamefont {Bancal}}, \
  and\ \bibinfo {author} {\bibfnamefont {Valerio}\ \bibnamefont {Scarani}},\
  }\bibfield  {title} {\enquote {\bibinfo {title} {Quantum randomness
  extraction for various levels of characterization of the devices},}\ }\href
  {\doibase 10.1088/1751-8113/47/42/424028} {\bibfield  {journal} {\bibinfo
  {journal} {Journal of Physics A: Mathematical and Theoretical}\ }\textbf
  {\bibinfo {volume} {47}},\ \bibinfo {pages} {424028} (\bibinfo {year}
  {2014})}\BibitemShut {NoStop}%
\bibitem [{\citenamefont {Passaro}\ \emph {et~al.}(2015)\citenamefont
  {Passaro}, \citenamefont {Cavalcanti}, \citenamefont {Skrzypczyk},\ and\
  \citenamefont {Ac{\'{\i}}n}}]{Passaro2015}%
  \BibitemOpen
  \bibfield  {author} {\bibinfo {author} {\bibfnamefont {Elsa}\ \bibnamefont
  {Passaro}}, \bibinfo {author} {\bibfnamefont {Daniel}\ \bibnamefont
  {Cavalcanti}}, \bibinfo {author} {\bibfnamefont {Paul}\ \bibnamefont
  {Skrzypczyk}}, \ and\ \bibinfo {author} {\bibfnamefont {Antonio}\
  \bibnamefont {Ac{\'{\i}}n}},\ }\bibfield  {title} {\enquote {\bibinfo {title}
  {Optimal randomness certification in the quantum steering and
  prepare-and-measure scenarios},}\ }\href {\doibase
  10.1088/1367-2630/17/11/113010} {\bibfield  {journal} {\bibinfo  {journal}
  {New Journal of Physics}\ }\textbf {\bibinfo {volume} {17}},\ \bibinfo
  {pages} {113010} (\bibinfo {year} {2015})}\BibitemShut {NoStop}%
\bibitem [{\citenamefont {Skrzypczyk}\ and\ \citenamefont
  {Cavalcanti}(2018)}]{Skrz2018}%
  \BibitemOpen
  \bibfield  {author} {\bibinfo {author} {\bibfnamefont {P.}~\bibnamefont
  {Skrzypczyk}}\ and\ \bibinfo {author} {\bibfnamefont {D.}~\bibnamefont
  {Cavalcanti}},\ }\bibfield  {title} {\enquote {\bibinfo {title} {Maximal
  randomness generation from steering inequality violations using qudits},}\
  }\href {\doibase 10.1103/PhysRevLett.120.260401} {\bibfield  {journal}
  {\bibinfo  {journal} {Phys. Rev. Lett.}\ }\textbf {\bibinfo {volume} {120}},\
  \bibinfo {pages} {260401} (\bibinfo {year} {2018})}\BibitemShut {NoStop}%
\bibitem [{\citenamefont {Gehring}\ \emph {et~al.}(2015)\citenamefont
  {Gehring}, \citenamefont {Händchen}, \citenamefont {Duhme}, \citenamefont
  {Furrer}, \citenamefont {Franz}, \citenamefont {Pacher}, \citenamefont
  {Werner},\ and\ \citenamefont {Schnabel}}]{Gehring2015}%
  \BibitemOpen
  \bibfield  {author} {\bibinfo {author} {\bibfnamefont {T.}~\bibnamefont
  {Gehring}}, \bibinfo {author} {\bibfnamefont {V.}~\bibnamefont {Händchen}},
  \bibinfo {author} {\bibfnamefont {J.}~\bibnamefont {Duhme}}, \bibinfo
  {author} {\bibfnamefont {F.}~\bibnamefont {Furrer}}, \bibinfo {author}
  {\bibfnamefont {T.}~\bibnamefont {Franz}}, \bibinfo {author} {\bibfnamefont
  {C.}~\bibnamefont {Pacher}}, \bibinfo {author} {\bibfnamefont {R.~F.}\
  \bibnamefont {Werner}}, \ and\ \bibinfo {author} {\bibfnamefont
  {R.}~\bibnamefont {Schnabel}},\ }\bibfield  {title} {\enquote {\bibinfo
  {title} {Implementation of continuous-variable quantum key distribution with
  composable and one-sided-device-independent security against coherent
  attacks},}\ }\href {\doibase 10.1038/ncomms9795} {\bibfield  {journal}
  {\bibinfo  {journal} {Nature Communications}\ }\textbf {\bibinfo {volume}
  {6}},\ \bibinfo {pages} {8795} (\bibinfo {year} {2015})}\BibitemShut
  {NoStop}%
\bibitem [{\citenamefont {Walk}\ \emph {et~al.}(2016)\citenamefont {Walk},
  \citenamefont {Hosseini}, \citenamefont {Geng}, \citenamefont {Thearle},
  \citenamefont {Haw}, \citenamefont {Armstrong}, \citenamefont {Assad},
  \citenamefont {Janousek}, \citenamefont {Ralph}, \citenamefont {Symul},
  \citenamefont {Wiseman},\ and\ \citenamefont {Lam}}]{Walk2016}%
  \BibitemOpen
  \bibfield  {author} {\bibinfo {author} {\bibfnamefont {N.}~\bibnamefont
  {Walk}}, \bibinfo {author} {\bibfnamefont {S.}~\bibnamefont {Hosseini}},
  \bibinfo {author} {\bibfnamefont {J.}~\bibnamefont {Geng}}, \bibinfo {author}
  {\bibfnamefont {O.}~\bibnamefont {Thearle}}, \bibinfo {author} {\bibfnamefont
  {J.~Y.}\ \bibnamefont {Haw}}, \bibinfo {author} {\bibfnamefont
  {S.}~\bibnamefont {Armstrong}}, \bibinfo {author} {\bibfnamefont {S.~M.}\
  \bibnamefont {Assad}}, \bibinfo {author} {\bibfnamefont {J.}~\bibnamefont
  {Janousek}}, \bibinfo {author} {\bibfnamefont {T.~C.}\ \bibnamefont {Ralph}},
  \bibinfo {author} {\bibfnamefont {T.}~\bibnamefont {Symul}}, \bibinfo
  {author} {\bibfnamefont {H.~M.}\ \bibnamefont {Wiseman}}, \ and\ \bibinfo
  {author} {\bibfnamefont {P.~K.}\ \bibnamefont {Lam}},\ }\bibfield  {title}
  {\enquote {\bibinfo {title} {Experimental demonstration of gaussian protocols
  for one-sided device-independent quantum key distribution},}\ }\href
  {\doibase 10.1364/OPTICA.3.000634} {\bibfield  {journal} {\bibinfo  {journal}
  {Optica}\ }\textbf {\bibinfo {volume} {3}},\ \bibinfo {pages} {634--642}
  (\bibinfo {year} {2016})}\BibitemShut {NoStop}%
\bibitem [{\citenamefont {M{\'{a}}ttar}\ \emph {et~al.}(2017)\citenamefont
  {M{\'{a}}ttar}, \citenamefont {Skrzypczyk}, \citenamefont {Aguilar},
  \citenamefont {Nery}, \citenamefont {Ribeiro}, \citenamefont {Walborn},\ and\
  \citenamefont {Cavalcanti}}]{Mattar2017}%
  \BibitemOpen
  \bibfield  {author} {\bibinfo {author} {\bibfnamefont {A}~\bibnamefont
  {M{\'{a}}ttar}}, \bibinfo {author} {\bibfnamefont {P}~\bibnamefont
  {Skrzypczyk}}, \bibinfo {author} {\bibfnamefont {G~H}\ \bibnamefont
  {Aguilar}}, \bibinfo {author} {\bibfnamefont {R~V}\ \bibnamefont {Nery}},
  \bibinfo {author} {\bibfnamefont {P~H~Souto}\ \bibnamefont {Ribeiro}},
  \bibinfo {author} {\bibfnamefont {S~P}\ \bibnamefont {Walborn}}, \ and\
  \bibinfo {author} {\bibfnamefont {D}~\bibnamefont {Cavalcanti}},\ }\bibfield
  {title} {\enquote {\bibinfo {title} {Experimental multipartite entanglement
  and randomness certification of the {W} state in the quantum steering
  scenario},}\ }\href {\doibase 10.1088/2058-9565/aa629b} {\bibfield  {journal}
  {\bibinfo  {journal} {Quantum Science and Technology}\ }\textbf {\bibinfo
  {volume} {2}},\ \bibinfo {pages} {015011} (\bibinfo {year}
  {2017})}\BibitemShut {NoStop}%
\bibitem [{\citenamefont {Wang}\ \emph {et~al.}(2018)\citenamefont {Wang},
  \citenamefont {Paesani}, \citenamefont {Ding}, \citenamefont {Santagati},
  \citenamefont {Skrzypczyk}, \citenamefont {Salavrakos}, \citenamefont {Tura},
  \citenamefont {Augusiak}, \citenamefont {Mančinska}, \citenamefont {Bacco},
  \citenamefont {Bonneau}, \citenamefont {Silverstone}, \citenamefont {Gong},
  \citenamefont {Acín}, \citenamefont {Rottwitt}, \citenamefont {Oxenløwe},
  \citenamefont {O’Brien}, \citenamefont {Laing},\ and\ \citenamefont
  {Thompson}}]{Wang2018}%
  \BibitemOpen
  \bibfield  {author} {\bibinfo {author} {\bibfnamefont {J.}~\bibnamefont
  {Wang}}, \bibinfo {author} {\bibfnamefont {S.}~\bibnamefont {Paesani}},
  \bibinfo {author} {\bibfnamefont {Y.}~\bibnamefont {Ding}}, \bibinfo {author}
  {\bibfnamefont {R.}~\bibnamefont {Santagati}}, \bibinfo {author}
  {\bibfnamefont {P.}~\bibnamefont {Skrzypczyk}}, \bibinfo {author}
  {\bibfnamefont {A.}~\bibnamefont {Salavrakos}}, \bibinfo {author}
  {\bibfnamefont {J.}~\bibnamefont {Tura}}, \bibinfo {author} {\bibfnamefont
  {R.}~\bibnamefont {Augusiak}}, \bibinfo {author} {\bibfnamefont
  {L.}~\bibnamefont {Mančinska}}, \bibinfo {author} {\bibfnamefont
  {D.}~\bibnamefont {Bacco}}, \bibinfo {author} {\bibfnamefont
  {D.}~\bibnamefont {Bonneau}}, \bibinfo {author} {\bibfnamefont {J.~W.}\
  \bibnamefont {Silverstone}}, \bibinfo {author} {\bibfnamefont
  {Q.}~\bibnamefont {Gong}}, \bibinfo {author} {\bibfnamefont {A.}~\bibnamefont
  {Acín}}, \bibinfo {author} {\bibfnamefont {K.}~\bibnamefont {Rottwitt}},
  \bibinfo {author} {\bibfnamefont {L.~K.}\ \bibnamefont {Oxenløwe}}, \bibinfo
  {author} {\bibfnamefont {J.~L.}\ \bibnamefont {O’Brien}}, \bibinfo {author}
  {\bibfnamefont {A.}~\bibnamefont {Laing}}, \ and\ \bibinfo {author}
  {\bibfnamefont {M.~G.}\ \bibnamefont {Thompson}},\ }\bibfield  {title}
  {\enquote {\bibinfo {title} {Multidimensional quantum entanglement with
  large-scale integrated optics},}\ }\href {\doibase 10.1126/science.aar7053}
  {\bibfield  {journal} {\bibinfo  {journal} {Science}\ }\textbf {\bibinfo
  {volume} {360}},\ \bibinfo {pages} {285--291} (\bibinfo {year}
  {2018})}\BibitemShut {NoStop}%
\bibitem [{\citenamefont {Larsen}\ \emph {et~al.}(2019)\citenamefont {Larsen},
  \citenamefont {Guo}, \citenamefont {Breum}, \citenamefont
  {Neergaard-Nielsen},\ and\ \citenamefont {Andersen}}]{Larsen19}%
  \BibitemOpen
  \bibfield  {author} {\bibinfo {author} {\bibfnamefont {M.~V.}\ \bibnamefont
  {Larsen}}, \bibinfo {author} {\bibfnamefont {X.}~\bibnamefont {Guo}},
  \bibinfo {author} {\bibfnamefont {C.~R.}\ \bibnamefont {Breum}}, \bibinfo
  {author} {\bibfnamefont {J.~S.}\ \bibnamefont {Neergaard-Nielsen}}, \ and\
  \bibinfo {author} {\bibfnamefont {U.~L.}\ \bibnamefont {Andersen}},\
  }\bibfield  {title} {\enquote {\bibinfo {title} {Fiber-coupled epr-state
  generation using a single temporally multiplexed squeezed light source},}\
  }\href {\doibase 10.1038/s41534-019-0170-y} {\bibfield  {journal} {\bibinfo
  {journal} {npj Quantum Inf}\ }\textbf {\bibinfo {volume} {5}},\ \bibinfo
  {pages} {46} (\bibinfo {year} {2019})}\BibitemShut {NoStop}%
\bibitem [{\citenamefont {Kashiwazaki}\ \emph {et~al.}(2020)\citenamefont
  {Kashiwazaki}, \citenamefont {Takanashi}, \citenamefont {Yamashima},
  \citenamefont {Kazama}, \citenamefont {Enbutsu}, \citenamefont {Kasahara},
  \citenamefont {Umeki},\ and\ \citenamefont {Furusawa}}]{Kashiwazaki2020}%
  \BibitemOpen
  \bibfield  {author} {\bibinfo {author} {\bibfnamefont {Takahiro}\
  \bibnamefont {Kashiwazaki}}, \bibinfo {author} {\bibfnamefont {Naoto}\
  \bibnamefont {Takanashi}}, \bibinfo {author} {\bibfnamefont {Taichi}\
  \bibnamefont {Yamashima}}, \bibinfo {author} {\bibfnamefont {Takushi}\
  \bibnamefont {Kazama}}, \bibinfo {author} {\bibfnamefont {Koji}\ \bibnamefont
  {Enbutsu}}, \bibinfo {author} {\bibfnamefont {Ryoichi}\ \bibnamefont
  {Kasahara}}, \bibinfo {author} {\bibfnamefont {Takeshi}\ \bibnamefont
  {Umeki}}, \ and\ \bibinfo {author} {\bibfnamefont {Akira}\ \bibnamefont
  {Furusawa}},\ }\bibfield  {title} {\enquote {\bibinfo {title}
  {Continuous-wave 6-db-squeezed light with 2.5-thz-bandwidth from single-mode
  ppln waveguide},}\ }\href {\doibase 10.1063/1.5142437} {\bibfield  {journal}
  {\bibinfo  {journal} {APL Photonics}\ }\textbf {\bibinfo {volume} {5}},\
  \bibinfo {pages} {036104} (\bibinfo {year} {2020})}\BibitemShut {NoStop}%
\bibitem [{\citenamefont {Fine}(1982)}]{Fine1982}%
  \BibitemOpen
  \bibfield  {author} {\bibinfo {author} {\bibfnamefont {A.}~\bibnamefont
  {Fine}},\ }\bibfield  {title} {\enquote {\bibinfo {title} {Hidden variables,
  joint probability, and the bell inequalities},}\ }\href {\doibase
  10.1103/PhysRevLett.48.291} {\bibfield  {journal} {\bibinfo  {journal} {Phys.
  Rev. Lett.}\ }\textbf {\bibinfo {volume} {48}},\ \bibinfo {pages} {291--295}
  (\bibinfo {year} {1982})}\BibitemShut {NoStop}%
\bibitem [{\citenamefont {Ou}\ \emph {et~al.}(1992)\citenamefont {Ou},
  \citenamefont {Pereira},\ and\ \citenamefont {Kimble}}]{Ou1992}%
  \BibitemOpen
  \bibfield  {author} {\bibinfo {author} {\bibfnamefont {Z.~Y.}\ \bibnamefont
  {Ou}}, \bibinfo {author} {\bibfnamefont {S.~F.}\ \bibnamefont {Pereira}}, \
  and\ \bibinfo {author} {\bibfnamefont {H.~J.}\ \bibnamefont {Kimble}},\
  }\bibfield  {title} {\enquote {\bibinfo {title} {Realization of the
  einstein-podolsky-rosen paradox for continuous variables in nondegenerate
  parametric amplification},}\ }\href {\doibase 10.1007/BF00325015} {\bibfield
  {journal} {\bibinfo  {journal} {Applied Physics B}\ }\textbf {\bibinfo
  {volume} {55}},\ \bibinfo {pages} {265--278} (\bibinfo {year}
  {1992})}\BibitemShut {NoStop}%
\bibitem [{\citenamefont {Händchen}\ \emph {et~al.}(2012)\citenamefont
  {Händchen}, \citenamefont {Eberle}, \citenamefont {Steinlechner},
  \citenamefont {Samblowski}, \citenamefont {Franz}, \citenamefont {Werner},\
  and\ \citenamefont {Schnabel}}]{Haendchen2012}%
  \BibitemOpen
  \bibfield  {author} {\bibinfo {author} {\bibfnamefont {V.}~\bibnamefont
  {Händchen}}, \bibinfo {author} {\bibfnamefont {T.}~\bibnamefont {Eberle}},
  \bibinfo {author} {\bibfnamefont {S.}~\bibnamefont {Steinlechner}}, \bibinfo
  {author} {\bibfnamefont {A.}~\bibnamefont {Samblowski}}, \bibinfo {author}
  {\bibfnamefont {T.}~\bibnamefont {Franz}}, \bibinfo {author} {\bibfnamefont
  {R.~F.}\ \bibnamefont {Werner}}, \ and\ \bibinfo {author} {\bibfnamefont
  {R.}~\bibnamefont {Schnabel}},\ }\bibfield  {title} {\enquote {\bibinfo
  {title} {Observation of one-way einstein-podolsky-rosen steering},}\ }\href
  {\doibase 10.1038/nphoton.2012.202} {\bibfield  {journal} {\bibinfo
  {journal} {Nature Photonics}\ }\textbf {\bibinfo {volume} {6}},\ \bibinfo
  {pages} {596--599} (\bibinfo {year} {2012})}\BibitemShut {NoStop}%
\bibitem [{\citenamefont {Armstrong}\ \emph {et~al.}(2015)\citenamefont
  {Armstrong}, \citenamefont {Wang}, \citenamefont {Teh}, \citenamefont {Gong},
  \citenamefont {He}, \citenamefont {Janousek}, \citenamefont {Bachor},
  \citenamefont {Reid},\ and\ \citenamefont {Lam}}]{Armstrong2015}%
  \BibitemOpen
  \bibfield  {author} {\bibinfo {author} {\bibfnamefont {S.}~\bibnamefont
  {Armstrong}}, \bibinfo {author} {\bibfnamefont {M.}~\bibnamefont {Wang}},
  \bibinfo {author} {\bibfnamefont {R.~Y.}\ \bibnamefont {Teh}}, \bibinfo
  {author} {\bibfnamefont {Q.}~\bibnamefont {Gong}}, \bibinfo {author}
  {\bibfnamefont {Q.}~\bibnamefont {He}}, \bibinfo {author} {\bibfnamefont
  {J.}~\bibnamefont {Janousek}}, \bibinfo {author} {\bibfnamefont {H.-A.}\
  \bibnamefont {Bachor}}, \bibinfo {author} {\bibfnamefont {M.~D.}\
  \bibnamefont {Reid}}, \ and\ \bibinfo {author} {\bibfnamefont {P.~K.}\
  \bibnamefont {Lam}},\ }\bibfield  {title} {\enquote {\bibinfo {title}
  {Multipartite einstein-podolsky-rosen steering and genuine tripartite
  entanglement with optical networks},}\ }\href {\doibase 10.1038/nphys3202}
  {\bibfield  {journal} {\bibinfo  {journal} {Nature Physics}\ }\textbf
  {\bibinfo {volume} {11}},\ \bibinfo {pages} {167--172} (\bibinfo {year}
  {2015})}\BibitemShut {NoStop}%
\bibitem [{\citenamefont {Deng}\ \emph {et~al.}(2017)\citenamefont {Deng},
  \citenamefont {Xiang}, \citenamefont {Tian}, \citenamefont {Adesso},
  \citenamefont {He}, \citenamefont {Gong}, \citenamefont {Su}, \citenamefont
  {Xie},\ and\ \citenamefont {Peng}}]{Deng2017}%
  \BibitemOpen
  \bibfield  {author} {\bibinfo {author} {\bibfnamefont {X.}~\bibnamefont
  {Deng}}, \bibinfo {author} {\bibfnamefont {Y.}~\bibnamefont {Xiang}},
  \bibinfo {author} {\bibfnamefont {C.}~\bibnamefont {Tian}}, \bibinfo {author}
  {\bibfnamefont {G.}~\bibnamefont {Adesso}}, \bibinfo {author} {\bibfnamefont
  {Q.}~\bibnamefont {He}}, \bibinfo {author} {\bibfnamefont {Q.}~\bibnamefont
  {Gong}}, \bibinfo {author} {\bibfnamefont {X.}~\bibnamefont {Su}}, \bibinfo
  {author} {\bibfnamefont {C.}~\bibnamefont {Xie}}, \ and\ \bibinfo {author}
  {\bibfnamefont {K.}~\bibnamefont {Peng}},\ }\bibfield  {title} {\enquote
  {\bibinfo {title} {Demonstration of monogamy relations for
  einstein-podolsky-rosen steering in gaussian cluster states},}\ }\href
  {\doibase 10.1103/PhysRevLett.118.230501} {\bibfield  {journal} {\bibinfo
  {journal} {Phys. Rev. Lett.}\ }\textbf {\bibinfo {volume} {118}},\ \bibinfo
  {pages} {230501} (\bibinfo {year} {2017})}\BibitemShut {NoStop}%
\bibitem [{\citenamefont {Qin}\ \emph {et~al.}(2017)\citenamefont {Qin},
  \citenamefont {Deng}, \citenamefont {Tian}, \citenamefont {Wang},
  \citenamefont {Su}, \citenamefont {Xie},\ and\ \citenamefont
  {Peng}}]{Qin2017}%
  \BibitemOpen
  \bibfield  {author} {\bibinfo {author} {\bibfnamefont {Z.}~\bibnamefont
  {Qin}}, \bibinfo {author} {\bibfnamefont {X.}~\bibnamefont {Deng}}, \bibinfo
  {author} {\bibfnamefont {C.}~\bibnamefont {Tian}}, \bibinfo {author}
  {\bibfnamefont {M.}~\bibnamefont {Wang}}, \bibinfo {author} {\bibfnamefont
  {X.}~\bibnamefont {Su}}, \bibinfo {author} {\bibfnamefont {C.}~\bibnamefont
  {Xie}}, \ and\ \bibinfo {author} {\bibfnamefont {K.}~\bibnamefont {Peng}},\
  }\bibfield  {title} {\enquote {\bibinfo {title} {Manipulating the direction
  of einstein-podolsky-rosen steering},}\ }\href {\doibase
  10.1103/PhysRevA.95.052114} {\bibfield  {journal} {\bibinfo  {journal} {Phys.
  Rev. A}\ }\textbf {\bibinfo {volume} {95}},\ \bibinfo {pages} {052114}
  (\bibinfo {year} {2017})}\BibitemShut {NoStop}%
\bibitem [{\citenamefont {Wang}\ \emph {et~al.}(2020)\citenamefont {Wang},
  \citenamefont {Xiang}, \citenamefont {Kang}, \citenamefont {Han},
  \citenamefont {Liu}, \citenamefont {He}, \citenamefont {Gong}, \citenamefont
  {Su},\ and\ \citenamefont {Peng}}]{Wang2020}%
  \BibitemOpen
  \bibfield  {author} {\bibinfo {author} {\bibfnamefont {M.}~\bibnamefont
  {Wang}}, \bibinfo {author} {\bibfnamefont {Y.}~\bibnamefont {Xiang}},
  \bibinfo {author} {\bibfnamefont {H.}~\bibnamefont {Kang}}, \bibinfo {author}
  {\bibfnamefont {D.}~\bibnamefont {Han}}, \bibinfo {author} {\bibfnamefont
  {Y.}~\bibnamefont {Liu}}, \bibinfo {author} {\bibfnamefont {Q.}~\bibnamefont
  {He}}, \bibinfo {author} {\bibfnamefont {Q.}~\bibnamefont {Gong}}, \bibinfo
  {author} {\bibfnamefont {X.}~\bibnamefont {Su}}, \ and\ \bibinfo {author}
  {\bibfnamefont {K.}~\bibnamefont {Peng}},\ }\bibfield  {title} {\enquote
  {\bibinfo {title} {Deterministic distribution of multipartite entanglement
  and steering in a quantum network by separable states},}\ }\href {\doibase
  10.1103/PhysRevLett.125.260506} {\bibfield  {journal} {\bibinfo  {journal}
  {Phys. Rev. Lett.}\ }\textbf {\bibinfo {volume} {125}},\ \bibinfo {pages}
  {260506} (\bibinfo {year} {2020})}\BibitemShut {NoStop}%
\bibitem [{\citenamefont {Cavalcanti}\ and\ \citenamefont
  {Skrzypczyk}(2016)}]{Cavalcanti2016}%
  \BibitemOpen
  \bibfield  {author} {\bibinfo {author} {\bibfnamefont {D}~\bibnamefont
  {Cavalcanti}}\ and\ \bibinfo {author} {\bibfnamefont {P}~\bibnamefont
  {Skrzypczyk}},\ }\bibfield  {title} {\enquote {\bibinfo {title} {Quantum
  steering: a review with focus on semidefinite programming},}\ }\href
  {\doibase 10.1088/1361-6633/80/2/024001} {\bibfield  {journal} {\bibinfo
  {journal} {Reports on Progress in Physics}\ }\textbf {\bibinfo {volume}
  {80}},\ \bibinfo {pages} {024001} (\bibinfo {year} {2016})}\BibitemShut
  {NoStop}%
\bibitem [{\citenamefont {Tasca}\ \emph {et~al.}(2018)\citenamefont {Tasca},
  \citenamefont {S\'anchez}, \citenamefont {Walborn},\ and\ \citenamefont
  {Rudnicki}}]{Tasca2018}%
  \BibitemOpen
  \bibfield  {author} {\bibinfo {author} {\bibfnamefont {D.~S.}\ \bibnamefont
  {Tasca}}, \bibinfo {author} {\bibfnamefont {P.}~\bibnamefont {S\'anchez}},
  \bibinfo {author} {\bibfnamefont {S.~P.}\ \bibnamefont {Walborn}}, \ and\
  \bibinfo {author} {\bibfnamefont {\L{}.}\ \bibnamefont {Rudnicki}},\
  }\bibfield  {title} {\enquote {\bibinfo {title} {Mutual unbiasedness in
  coarse-grained continuous variables},}\ }\href {\doibase
  10.1103/PhysRevLett.120.040403} {\bibfield  {journal} {\bibinfo  {journal}
  {Phys. Rev. Lett.}\ }\textbf {\bibinfo {volume} {120}},\ \bibinfo {pages}
  {040403} (\bibinfo {year} {2018})}\BibitemShut {NoStop}%
\bibitem [{\citenamefont {Impagliazzo}\ \emph {et~al.}(1989)\citenamefont
  {Impagliazzo}, \citenamefont {Levin},\ and\ \citenamefont
  {Luby}}]{Impagliazzo1989}%
  \BibitemOpen
  \bibfield  {author} {\bibinfo {author} {\bibfnamefont {R.}~\bibnamefont
  {Impagliazzo}}, \bibinfo {author} {\bibfnamefont {L.~A.}\ \bibnamefont
  {Levin}}, \ and\ \bibinfo {author} {\bibfnamefont {M.}~\bibnamefont {Luby}},\
  }\bibfield  {title} {\enquote {\bibinfo {title} {Pseudo-random generation
  from one-way functions},}\ \ }(\bibinfo  {publisher} {Association for
  Computing Machinery},\ \bibinfo {address} {New York, NY, USA},\ \bibinfo
  {year} {1989})\ p.\ \bibinfo {pages} {12–24}\BibitemShut {NoStop}%
\bibitem [{\citenamefont {Boyd}\ and\ \citenamefont
  {Vandenberghe}(2004)}]{Boyd2004}%
  \BibitemOpen
  \bibfield  {author} {\bibinfo {author} {\bibfnamefont {Stephen}\ \bibnamefont
  {Boyd}}\ and\ \bibinfo {author} {\bibfnamefont {Lieven}\ \bibnamefont
  {Vandenberghe}},\ }\href@noop {} {\emph {\bibinfo {title} {Convex
  optimization}}}\ (\bibinfo  {publisher} {Cambridge university press},\
  \bibinfo {year} {2004})\BibitemShut {NoStop}%
\bibitem [{\citenamefont {Joch}\ \emph {et~al.}(2021)\citenamefont {Joch},
  \citenamefont {Slussarenko}, \citenamefont {Wang}, \citenamefont {Pepper},
  \citenamefont {Xie}, \citenamefont {Xu}, \citenamefont {Berkman},
  \citenamefont {Rogge},\ and\ \citenamefont {Pryde}}]{Joch2021}%
  \BibitemOpen
  \bibfield  {author} {\bibinfo {author} {\bibfnamefont {D.~J.}\ \bibnamefont
  {Joch}}, \bibinfo {author} {\bibfnamefont {S.}~\bibnamefont {Slussarenko}},
  \bibinfo {author} {\bibfnamefont {Y.}~\bibnamefont {Wang}}, \bibinfo {author}
  {\bibfnamefont {A.}~\bibnamefont {Pepper}}, \bibinfo {author} {\bibfnamefont
  {S.}~\bibnamefont {Xie}}, \bibinfo {author} {\bibfnamefont {B.-B.}\
  \bibnamefont {Xu}}, \bibinfo {author} {\bibfnamefont {I.~R.}\ \bibnamefont
  {Berkman}}, \bibinfo {author} {\bibfnamefont {S.}~\bibnamefont {Rogge}}, \
  and\ \bibinfo {author} {\bibfnamefont {G.~J.}\ \bibnamefont {Pryde}},\
  }\bibfield  {title} {\enquote {\bibinfo {title} {Certified random number
  generation from quantum steering},}\ }\href
  {https://export.arxiv.org/abs/2111.09506} {\bibfield  {journal} {\bibinfo
  {journal} {arXiv:2111.09506 [quant-ph]}\ } (\bibinfo {year}
  {2021})}\BibitemShut {NoStop}%
\bibitem [{\citenamefont {Dodonov}\ \emph {et~al.}(1994)\citenamefont
  {Dodonov}, \citenamefont {Man'ko},\ and\ \citenamefont {Man'ko}}]{Dodonov94}%
  \BibitemOpen
  \bibfield  {author} {\bibinfo {author} {\bibfnamefont {V.~V.}\ \bibnamefont
  {Dodonov}}, \bibinfo {author} {\bibfnamefont {O.~V.}\ \bibnamefont {Man'ko}},
  \ and\ \bibinfo {author} {\bibfnamefont {V.~I.}\ \bibnamefont {Man'ko}},\
  }\bibfield  {title} {\enquote {\bibinfo {title} {Multidimensional hermite
  polynomials and photon distribution for polymode mixed light},}\ }\href
  {\doibase 10.1103/PhysRevA.50.813} {\bibfield  {journal} {\bibinfo  {journal}
  {Phys. Rev. A}\ }\textbf {\bibinfo {volume} {50}},\ \bibinfo {pages} {813}
  (\bibinfo {year} {1994})}\BibitemShut {NoStop}%
\bibitem [{\citenamefont {Kok}\ and\ \citenamefont {Braunstein}(2001)}]{Kok01}%
  \BibitemOpen
  \bibfield  {author} {\bibinfo {author} {\bibfnamefont {P.}~\bibnamefont
  {Kok}}\ and\ \bibinfo {author} {\bibfnamefont {S.~L.}\ \bibnamefont
  {Braunstein}},\ }\bibfield  {title} {\enquote {\bibinfo {title}
  {Multi-dimensional hermite polynomials in quantum optics},}\ }\href {\doibase
  10.1088/0305-4470/34/31/312} {\bibfield  {journal} {\bibinfo  {journal} {J.
  Phys. A: Math. Gen.}\ }\textbf {\bibinfo {volume} {34}},\ \bibinfo {pages}
  {6185} (\bibinfo {year} {2001})}\BibitemShut {NoStop}%
\bibitem [{\citenamefont {Larsen}\ \emph {et~al.}(2021)\citenamefont {Larsen},
  \citenamefont {Guo}, \citenamefont {Breum}, \citenamefont
  {Neergaard-Nielsen},\ and\ \citenamefont {Andersen}}]{Larsen21}%
  \BibitemOpen
  \bibfield  {author} {\bibinfo {author} {\bibfnamefont {M.~V.}\ \bibnamefont
  {Larsen}}, \bibinfo {author} {\bibfnamefont {X.}~\bibnamefont {Guo}},
  \bibinfo {author} {\bibfnamefont {C.~R.}\ \bibnamefont {Breum}}, \bibinfo
  {author} {\bibfnamefont {J.~S.}\ \bibnamefont {Neergaard-Nielsen}}, \ and\
  \bibinfo {author} {\bibfnamefont {U.~L.}\ \bibnamefont {Andersen}},\
  }\bibfield  {title} {\enquote {\bibinfo {title} {Deterministic multi-mode
  gates on a scalable photonic quantum computing platform},}\ }\href {\doibase
  10.1038/s41567-021-01296-y} {\bibfield  {journal} {\bibinfo  {journal}
  {Nature Physics}\ }\textbf {\bibinfo {volume} {17}},\ \bibinfo {pages}
  {1018–1023} (\bibinfo {year} {2021})}\BibitemShut {NoStop}%
\end{thebibliography}%

\appendix

\section{Gaussian states with noise and loss}
\label{appendix:gaussian states}
Consider a continuous-variable system of $d$ bosonic modes. Associated with each mode is a pair of creation and annihilation operators that satisfy the canonical commutation relations $[\hat{a}_j,\hat{a}_k^\dag] = \delta_{jk}$, $[\hat{a}_j,\hat{a}_k]=0$ and $[\hat{a}^\dag_j,\hat{a}^\dag_k]=0$. The corresponding quadrature operators for each mode are defined as
\begin{equation}
    \hat{q}_j = \frac{1}{\sqrt{2}}\left(\hat{a}_j^\dag + \hat{a}_j\right), \quad \hat{p}_j = \frac{i}{\sqrt{2}}\left(\hat{a}_j^\dag - \hat{a}_j\right),
\end{equation}
and fulfil the commutation relations $[\hat{q}_j,\hat{p}_k] = i\delta_{jk}$. By definition, a Gaussian state has a Wigner function of the form
\begin{equation}
\label{eqn:gaussian state}
W(z) = \frac{\sqrt{\det G}}{\pi^d}\ue^{-(z-Z)\cdot G(z-Z)},
\end{equation}
where $z = (q,p) \in \R^d \times \R^d$ are canonical phase-space coordinates, $Z$ is a vector of expectation values $Z_k = \la \hat{z}_k\ra$ with $\hat z = (\hat{q}_1,\ldots,\hat{q}_d,\hat{p}_1,\ldots,\hat{p}_d)$, and $G$ is a real, symmetric and positive definite matrix. A Gaussian state is therefore completely characterised by the first moments $Z$ and the covariances of the quadrature operators
\begin{equation}
\label{eqn:covariances}
(G^{-1})_{jk} = \la \hat{z}_j \hat{z}_k + \hat{z}_k \hat{z}_j\ra - 2\la\hat{z}_j\ra\la\hat{z}_k\ra.
\end{equation}
The matrix elements of a Gaussian state in the Fock basis can be expressed in terms of $Z$, $G$ and multi-dimensional Hermite polynomials \cite{Dodonov94}, where the latter can be generated recursively \cite{Kok01}. The Gaussian states $\hat\rho$ in this work have first moments $Z$ equal to zero. In the following, we assume $Z=0$.

We model noise and loss by fictitious beam splitters with transmittivity $\eta$ between the source and each party. Thermal states of mean photon number $\bar{n}$ enter the other port of the beam splitters. The noise and losses are thus assumed to be symmetric for Alice and Bob. The case of pure loss is obtained by setting $\bar{n}=0$.

Before the beam splitters the total state is $\hat \rho \otimes \hat{\rho}_{th} \otimes \hat{\rho}_{th}$, where $\hat\rho$ is a two-mode Gaussian state produced by the source and each $\hat{\rho}_{th}$ is a thermal state with mean photon number $\bar{n}$. The corresponding Wigner function is $W_\rho(z)W_{th}(z_{t_1})W_{th}(z_{t_2})$, where $W_\rho(z)$ is a Gaussian with $G_\rho$ and $W_{th}(z_t)$ is a Gaussian with $G_{th} = (1+2\bar{n})^{-1}I.$ By combining the phase-space coordinates of the thermal states into $z_t = (q_t,p_t) \in \R^2 \times \R^2$, it can be shown that the beam splitters perform the transformations
\begin{align}
z & \to \sqrt{\eta}z + \sqrt{1-\eta}z_t,\label{eqn:bs 1}\\
z_t & \to \sqrt{\eta}z_t - \sqrt{1-\eta}z.\label{eqn:bs 2}
\end{align}
Integrating out the $z_t$ coordinates of the transformed state yields a Gaussian Wigner function with first moments equal to zero and
\begin{align}
\label{eqn:noisy G}
G &= \eta G_\rho + (1-\eta)G_t-\\& \eta(1-\eta)(G_t-G_\rho)\left[\eta G_t+ (1-\eta)G_\rho\right]^{-1} (G_t-G_\rho),\nonumber
\end{align}
where $G_t = G_{th}\oplus G_{th}$. This is the initial state with noise and loss applied.

The mean photon number is chosen such that the added noise corresponds to 1\% of the vacuum variance (i.e.~0.01 shot-noise units), which is the case when $\bar{n} = [200(1-\eta)]^{-1}$.

\section{Finite dimension}
\label{appendix:finite dimension}

 \begin{figure}
	\begin{center}
		\includegraphics[width=0.4\textwidth]{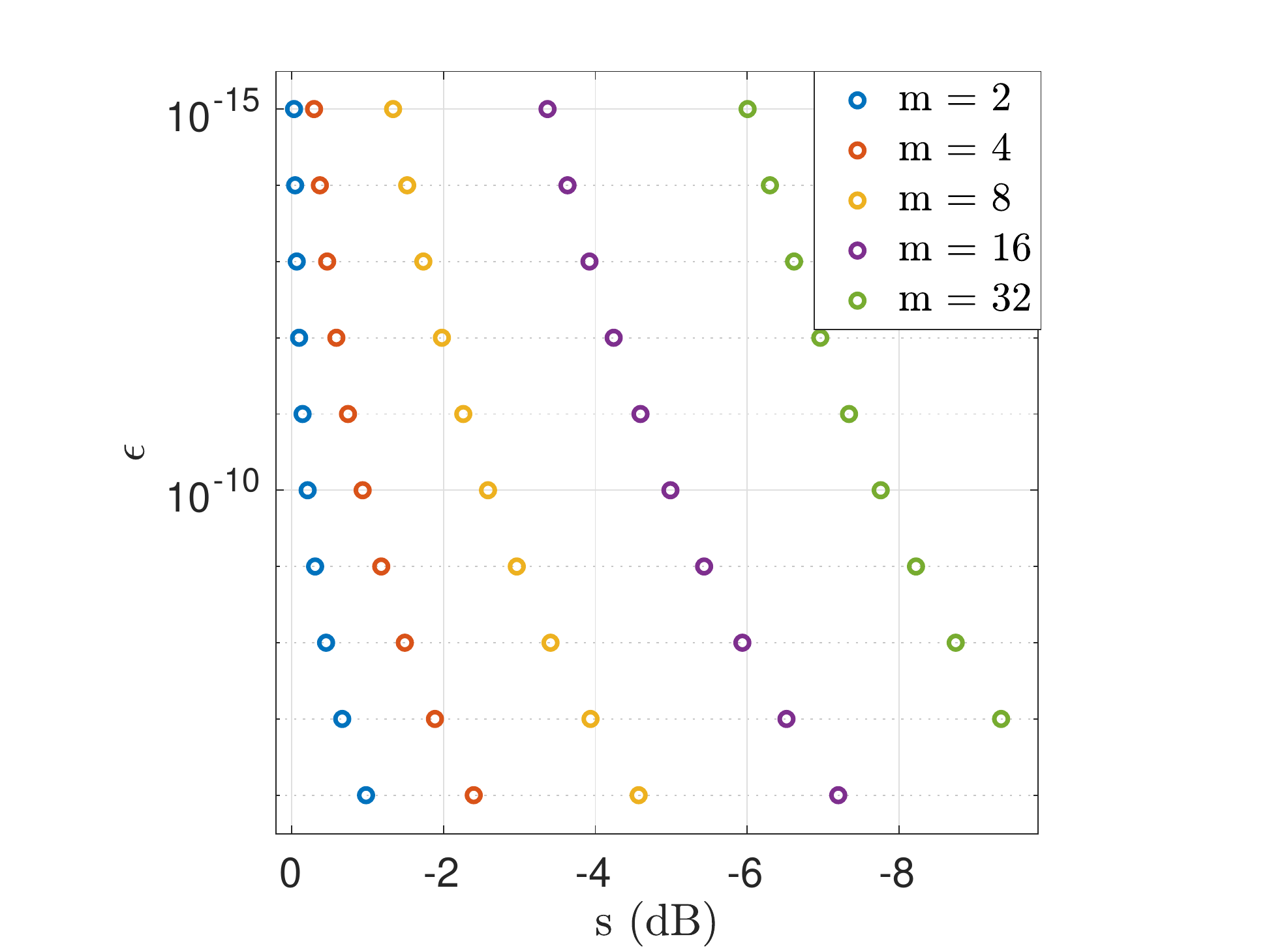}
	\end{center}
	\caption{\footnotesize Orthogonality parameter $\epsilon$ as a function of the squeezing parameter $s$ and the cut off in photon number $m$.}
	\label{fig:epsilon}
\end{figure}
 
 \begin{figure}
	\begin{center}
		\includegraphics[width=0.5\textwidth]{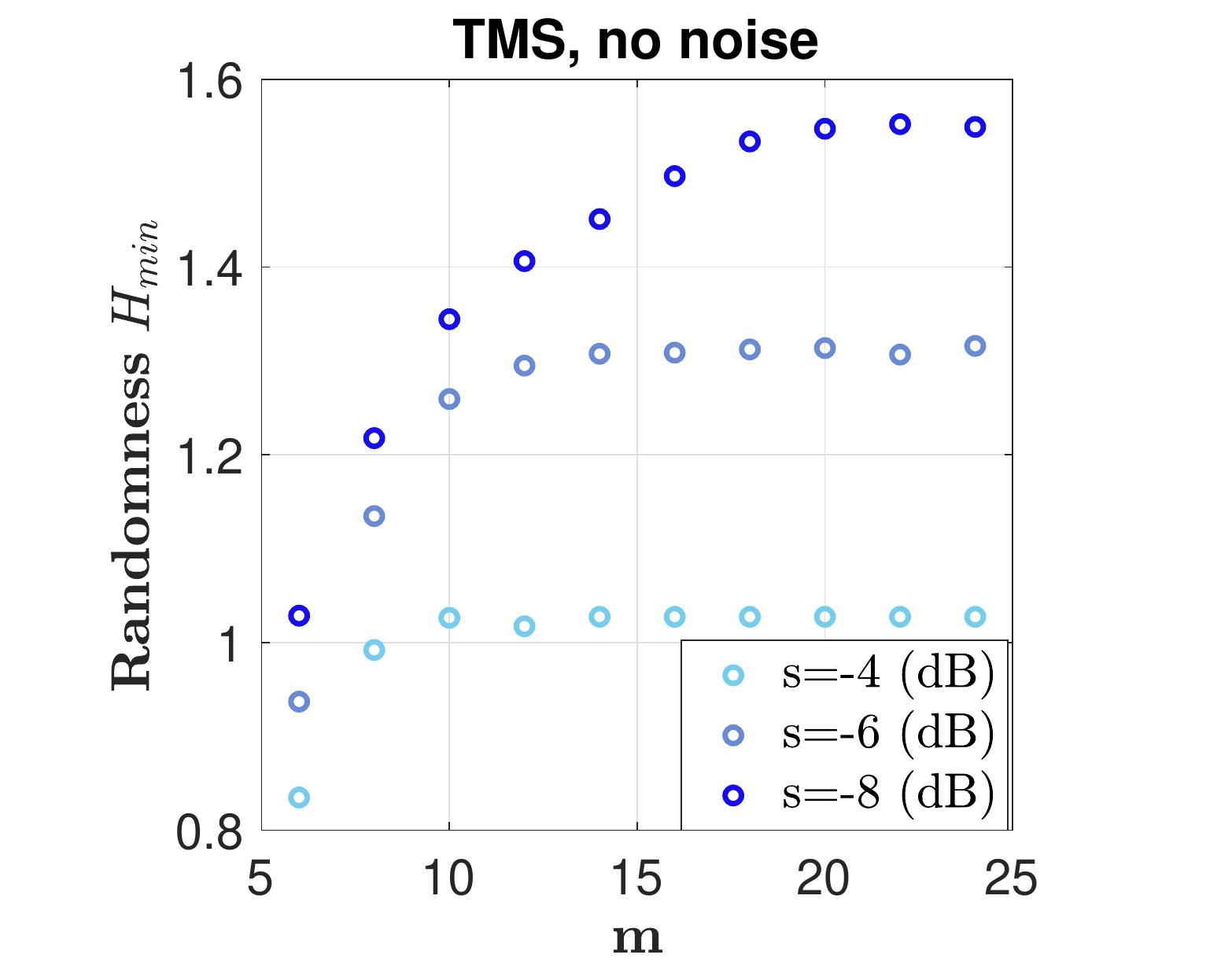}
	\end{center}
	\caption{\footnotesize Bound on the min-entropy $H_{\textnormal{min}}$ as a function of the cut off $m$ in the Fock basis for different squeezing parameters. Alice's measurement parameters are $o_A=8$ and $T_q=3$. }
	\label{fig:truncation}
\end{figure}

In order to numerically solve the SDPs we need to work in finite dimension. We use the Fock-space representation, and truncate both the two-mode squeezed vacuum (TMS) and the single-mode squeezed vacuum (SMS) states at a given photon number. In the following, we show that if this cut off is taken sufficiently large, then the min-entropy is unaffected.

The TMS state in the Fock basis is given by
\begin{equation}
    \ket{TMS} = \frac{1}{\cosh(\tilde{s})}\sum_{n=0}^\infty \tanh^{n}(\tilde{s})\ket{nn} ,
\end{equation}
where $\tilde{s}$ is the squeezing parameter, and $s$ expressed in dB is defined as $s=10\log_{10}(e^{-2\tilde{s}})$ [dB]. The normalised state after a cut off at $m$ photons has the form 
\begin{equation}
    \ket{TMS,m} = \sqrt{\frac{1-\tanh^2(\tilde{s})}{1-\tanh^{2(m+1)}(\tilde{s})}}\sum_{n=0}^m\tanh^n(\tilde{s})\ket{nn}.
\end{equation}
The deviation of the truncated state from the true state can be quantified in terms of the overlap $\braket{TMS}{TMS,m}=1-\epsilon$. \figref{fig:epsilon} illustrates the relation between the deviation $\epsilon$, the squeezing $s$ (dB) and the cut off $m$. 

It is computationally expensive to apply a large cut off. Indeed, in the SDP (2) of the main text, the dimension of the optimisation variables $\hat{\sigma}^e_{a|x}\in \mathbb{C}^{(m+1)\times(m+1)}$ and the positivity constraints (2d) are bottlenecks of the optimisation. Hence, it is desirable to keep $m$ as small as possible, while also minimising the error in computing $H_{\textnormal{min}}$. In ~\figref{fig:truncation} we observe that the numerical calculations of $H_{\textnormal{min}}$ stabilise at sufficiently high cut-off numbers. Note that while we considered the TMS state in these plots, the same behaviour holds for the SMS state.

\section{A lower bound on the min-entropy}
\label{appendix:lower bound}
Here we provide details on how to lower bound the min-entropy of the experimental data.

First we obtain an approximation of the initial Gaussian state using an idealised theoretical model of the experiment. To this end, we adapt the derivations in the supplementary information of Ref.~\cite{Larsen19} and Ref.~\cite{Larsen21} to find the quadrature squeezings
\begin{equation}
    \textnormal{Var}\left[\hat{z}\right] = \int_\R\int_\R f(t)f(t')\la \hat{z}(t)\hat{z}(t')\ra dt dt', \quad \hat{z}=\hat{q},\hat{p}.
\end{equation}
Here
\begin{equation}
    f(t) = \frac{1}{\sqrt{N}}\sin(\omega t)\ue^{-t^2/2\sigma^2}
\end{equation}
is the temporal mode function, chosen to optimize the measured squeezing, with $\sigma = 270 \, \textnormal{ns}$, $\omega = 2\pi \times 2.72 \, \textnormal{MHz}$, and $N$ is a normalisation factor defined by $\int_\R f^2(t) dt = 1$. The quadrature auto-covariance functions $\la \hat{z}(t)\hat{z}(t')\ra$ are given by
\begin{align}
    \la \hat{q}(t)\hat{q}(t')\ra &= \frac{1}{2}\delta(t-t') + \frac{\eta \gamma \nu}{\gamma-\nu}\ue^{-(\gamma-\nu)|t-t'|},\\
    \la \hat{p}(t)\hat{p}(t')\ra &= \frac{1}{2}\delta(t-t') - \frac{\eta \gamma \nu}{\gamma+\nu}\ue^{-(\gamma+\nu)|t-t'|},
\end{align}
where $\eta = 0.68$, $\gamma = 2\pi \times 8.1 \, \textnormal{MHz}$ and $\nu = 2\pi \times 5.2 \, \textnormal{MHz}$ are the overall efficiency, the OPO decay rate, and the pump rate respectively.

From this we are able to calculate the matrix 
\begin{equation}
    \label{eqn:experimental G}
    G = \begin{pmatrix} g_1 & g_2 & 0 & 0\\g_2 & g_1 & 0 & 0\\0 & 0 & g_1 & -g_2 \\ 0 & 0 & -g_2 & g_1 \end{pmatrix},
\end{equation}
with $g_1 = 1.38$ and $g_2 = 1.2597$, which completely characterises the Gaussian state (the first moments are zero). Using this state, together with the POVMs for Alice and Bob's measurements, the probability distributions $P_\textnormal{theory}(ab|xy)=\textnormal{tr}[\hat{M}_{a|x}\otimes\hat{M}_{b|y}\hat{\rho}_G]$ can be computed.

\begin{figure}
\footnotesize 
	\begin{center}
		\includegraphics[width=0.4\textwidth]{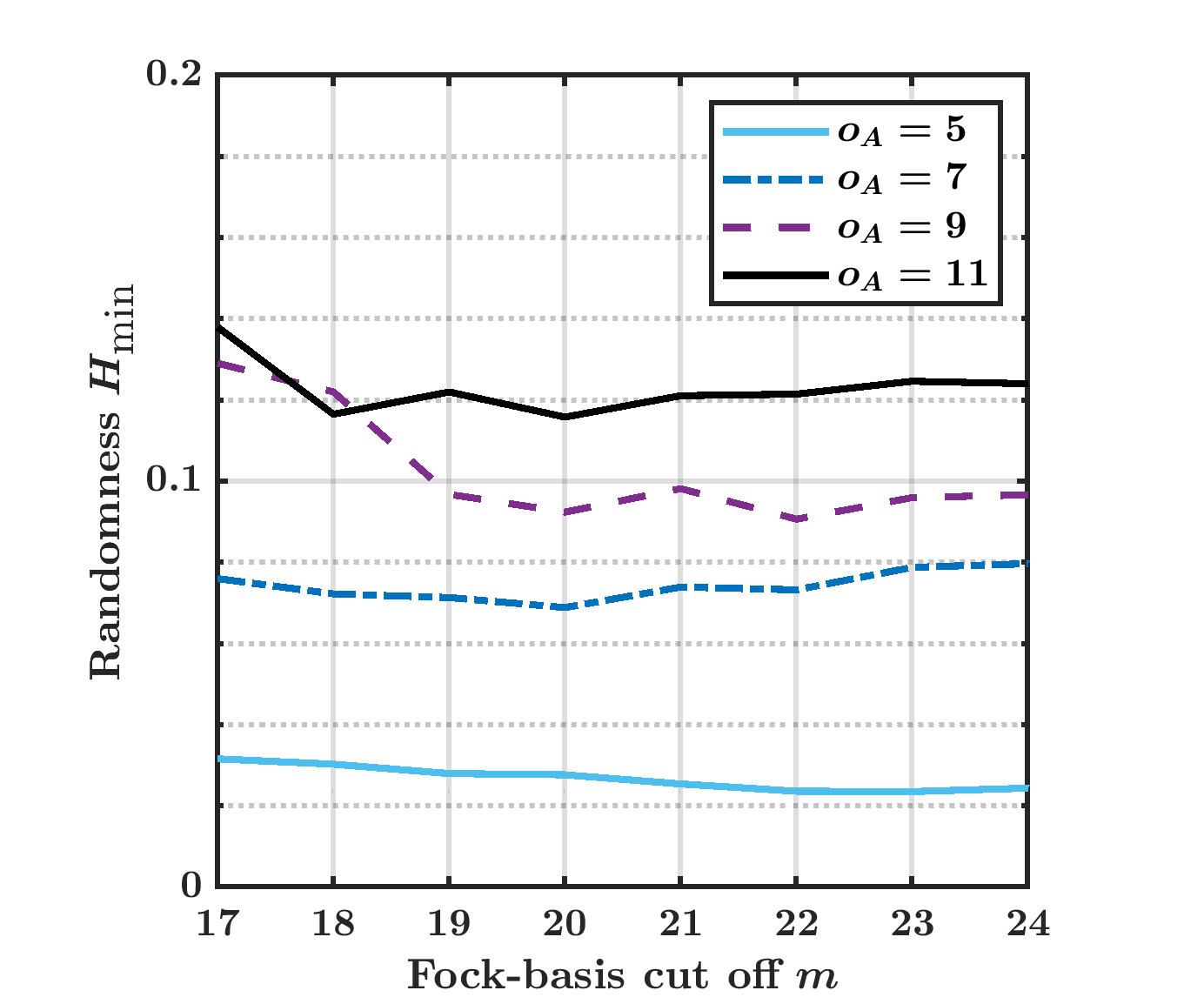}
	\end{center}
	\caption{\footnotesize Lower bound on the min-entropy vs Fock-basis cut off, for several values of Alice's outputs $o_A$. The measurement settings at each value of $o_A$ correspond to the maximum values of $H_\textnormal{min}$ in Fig. 3 of the main text.}
	\label{fig:hconvergence}
\end{figure}

Next we derive the dual of the SDP (2) in the main text. Recall that the condition (2d) must be replaced when Bob performs POVMs with the elements $\hat{M}_{b|y}$. We shall simply quote the result. However, a similar calculation can be found in Appendix C of \cite{Passaro2015}. In the dual formulation, given the data $P(ab|xy)$, the guessing probability can be computed via the following optimisation
\begin{align}
\min_{\{\xi_{abxy}\},\{G_x^e\}} \quad\!\! \sum_{a,b,x,y}\xi_{abxy} P(ab|xy),& \\
\textnormal{s.t.}\sum_{b,y}\xi_{abxy}\hat{M}_{b|y} - \delta_{ae}\delta_{xx^\ast}\hat{I} & \nonumber \\ +\delta_{xx^\ast}\sum_{x'}G_{x'}^e - G_x^e &\geq 0 \quad \forall a,e,x,
\end{align}
where $\xi_{abxy} \in \R$ and the $G_x^e$ are Hermitian matrices. Note that strong duality holds, and the optimal value of the dual is equal to the optimal value of the primal. After inserting the theoretical distributions $P_\textnormal{theory}(ab|xy)$ into the dual SDP, the resulting optimal dual variables $\xi_{abxy}$ can be used to obtain an upper bound on the guessing probability of the experimental data $u = \sum_{a,b,x,y}\xi_{abxy}P_{exp}(ab|xy)$. This then yields a lower bound on the min-entropy of the experimental data $h_l = -\log_2(u)$.

The Gaussian state (\ref{eqn:experimental G}) is quite spread out in phase space, therefore, we take the Fock-basis cut off to be $m = 24$. In Fig. \ref{fig:hconvergence} we illustrate that the lower bound on the min-entropy of the experimental data is fairly well converged with this choice. While there are some small fluctuations, there is very little change between the cut-off numbers $m=19$ and $m=24$. We are unable to go higher due to numerical limitations.  

\end{document}